\documentclass{aa}

\usepackage{amsmath,amssymb,amstext}

\usepackage[breaklinks,colorlinks,citecolor=blue,linkcolor=magenta]{hyperref}

\usepackage[all]{hypcap} 
\usepackage{multirow}
\usepackage{booktabs}
\usepackage{enumitem}
\usepackage{rotating}
\usepackage{xcolor}

\usepackage[percent]{overpic}
\usepackage{lineno}
\usepackage[T1]{fontenc}

\usepackage{graphicx}
\usepackage{txfonts}
\begin{document} 

   \title{The Smallest Scale of Hierarchy Survey (SSH)}
   \subtitle{III. Dwarf-dwarf satellite merging phenomena in the low-mass regime}

   \author{Elena Sacchi
          \inst{1}\fnmsep\inst{2}
          \and
          Michele Bellazzini\inst{2} \and
          Francesca Annibali\inst{2} \and 
          Monica Tosi\inst{2} \and
          Giacomo Beccari\inst{3} \and
          John M. Cannon\inst{4} \and
          Laura C. Hunter\inst{5} \and
          Diego Paris\inst{6} \and
          Sambit Roychowdhury\inst{7} \and 
          Lila Schisgal\inst{4} \and
          Liese van Zee\inst{8} \and
          Michele Cignoni\inst{2}\fnmsep\inst{9}\fnmsep\inst{10} \and
          Felice Cusano\inst{2} \and
          Roelof S. de Jong\inst{1} \and
          Leslie Hunt\inst{11} \and
          Raffaele Pascale\inst{2}
          }

   \institute{
   Leibniz-Institut für Astrophysik Potsdam (AIP), An der Sternwarte 16, 14482 Potsdam, Germany\\
   \email{esacchi@aip.de}
    \and
    INAF--Osservatorio di Astrofisica e Scienza dello Spazio di Bologna, Via Gobetti 93/3, I-40129 Bologna, Italy
    \and
    European Southern Observatory, Karl-Schwarzschild-Strasse 2, 85748, Garching bei München, Germany
    \and
    Macalester College, 1600 Grand Avenue, Saint Paul, MN 55105, USA
    \and
    Department of Physics and Astronomy, Dartmouth College, 17 Fayerweather Hill Rd, Hanover, NH 03755, USA
    \and
    INAF—Osservatorio Astronomico di Roma, Via Frascati 33, I-00078 Monte Porzio Catone, Italy
    \and
    University Observatory, Faculty of Physics, Ludwig-Maximilians-Universit\"{a}t, Scheinerstr. 1, 81679 M\"{u}nchen, Germany
    \and
    Department of Astronomy, Indiana University, 727 East 3rd Street, Bloomington, IN 47405, USA
    \and
    Department of Physics – University of Pisa, Largo B. Pontecorvo 3, 56127 Pisa, Italy
    \and
    INFN – Istituto Nazionale di Fisica Nucleare, Largo B. Pontecorvo 3, 56127 Pisa, Italy
    \and
    INAF – Osservatorio Astrofisico di Arcetri, Largo E. Fermi 5, 50125 Firenze, Italy 
             }

   \date{Received XXX; accepted XXX}

 
  \abstract
   {We present new deep, wide-field Large Binocular Telescope (LBT) $g$ and $r$ imaging data from the Smallest Scale of Hierarchy Survey (SSH) revealing previously undetected tidal features and stellar streams in the outskirts of six dwarf irregular galaxies (NGC~5238, UGC~6456, UGC~6541, UGC~7605, UGC~8638, and UGC~8760) with stellar masses in the range $1.2 \times 10^7$ M$_{\odot}$ to $1.4 \times 10^8$ M$_{\odot}$. The six dwarfs are located 1-2 Mpc away from large galaxies, implying that the observed distortions are unlikely to be due to tidal effects from a nearby, massive companion. At the dwarfs' distances of $\sim$3-4 Mpc, the identified tidal features are all resolved into individual stars in the LBT images and appear to be made of a population older than 1-2 Gyr, excluding the possibility that they result from irregular and asymmetric star formation episodes that are common in gas-rich dwarf galaxies. The most plausible explanation is that we are witnessing the hierarchical merging assembling of these dwarfs with their satellite populations, a scenario also supported by the peculiar morphology and disturbed velocity field of their HI component. From the SSH sample we estimate a fraction of late type dwarfs showing signs of merging with satellites of $\sim$13\%, in agreement with other recent independent studies and theoretical predictions within the $\Lambda$CDM cosmological framework.}


   \maketitle


\begin{figure*}
\centering
\includegraphics[width=\textwidth]{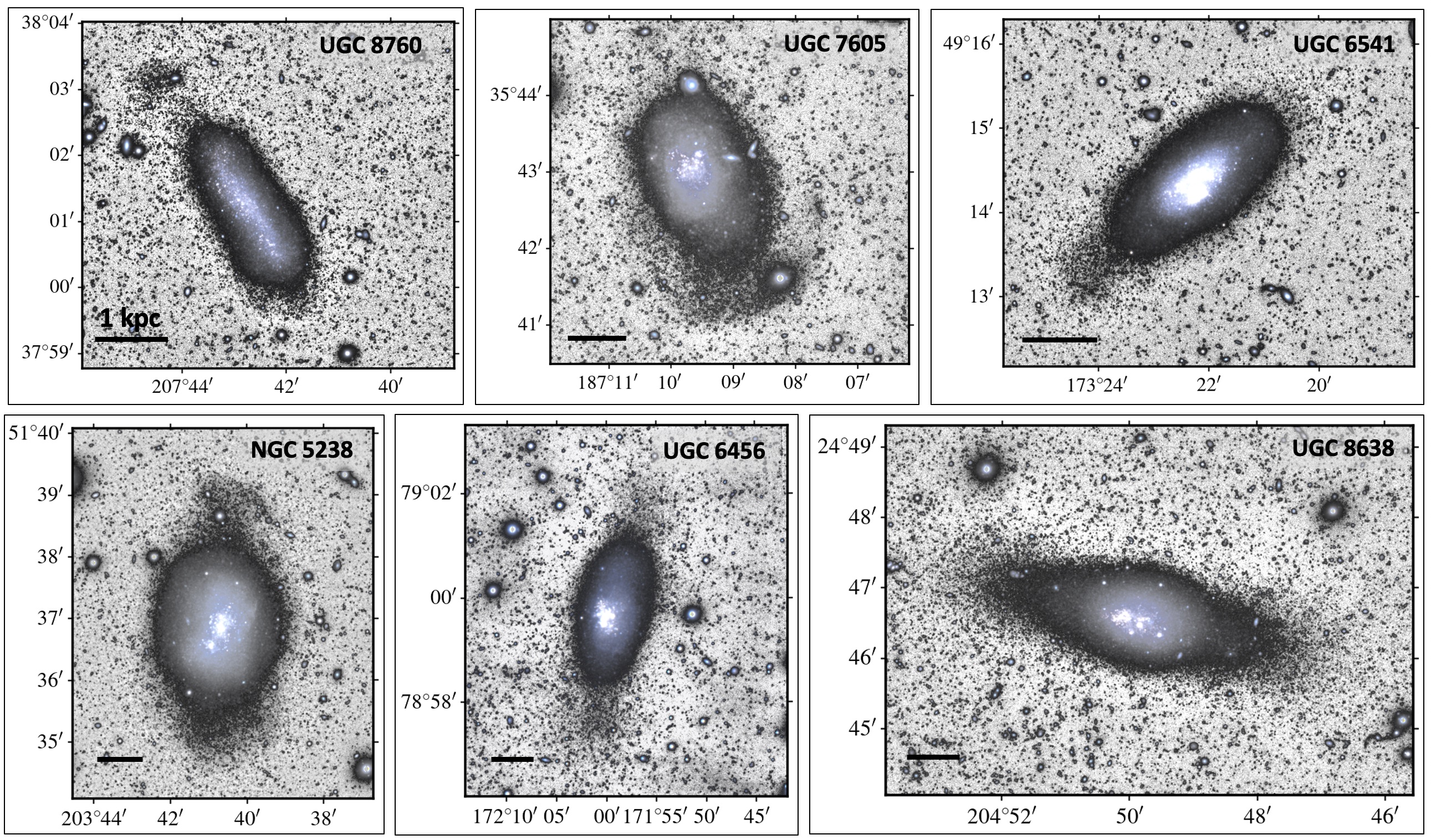}
\caption{LBT images of the six galaxies in the SSH survey exhibiting clear tidal features in their external regions. The grey-scale maps highlight the low surface brightness emission, while $g$, $r$ color-combined maps are used to emphasise the morphology of the central starburst regions. The black horizontal bar is 1 kpc long, adopting the distances listed in Table \ref{tab:table}. (North is up, East is left) 
}   
\label{fig:nicegal}
\end{figure*}

\begin{figure*}
\centering
\includegraphics[width=\linewidth]{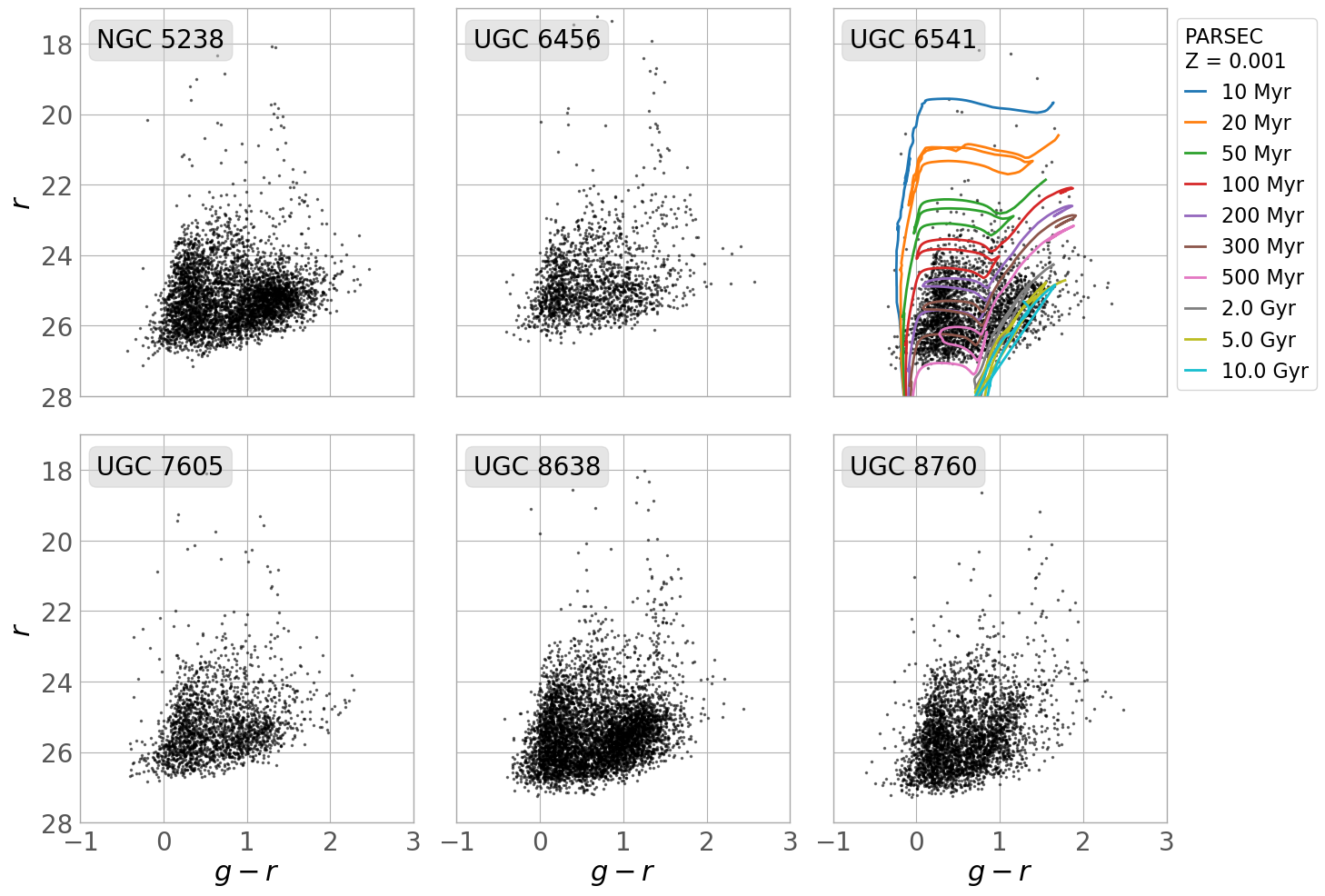}
\caption{$r$ vs $g-r$ CMDs from our new SSH data of the six galaxies analyzed here (see labels). As an example, overplotted on UGC 6541 are the PARSEC isochrones of different ages (as labelled) shifted to a distance of 3.89 Mpc and corrected for a foreground extinction $A_V = 0.052$; the isochrones’ metallicity of Z = 0.001 is consistent with the H II region oxygen abundance of 12 + log(O/H) = 7.82 \citep{Berg2012}.}
\label{fig:cmds-all}
\end{figure*}

\section{Introduction}
\setcounter{footnote}{11}
\urlstyle{sf}

The Lambda Cold Dark Matter ($\Lambda$CDM) cosmological scenario has proven remarkably successful at explaining the origin and evolution of large-scale structures in the Universe \citep{White78,Planck2016}. However, when zooming-in to the dwarf galaxy regime, the comparison between predictions and observations becomes less extensive and somewhat challenging \citep{Bullock2017}.
In particular, $\Lambda$CDM's self-similarity predicts that even low-mass host halos should have substructures down to the resolution limit of the simulations  \citep{Wetzel2016,Dooley2017,Besla2018,Jahn2019,Wang2020}, thus present-day dwarf galaxies should be surrounded by satellites and stellar streams \citep{Diemand2008}, a circumstance not sufficiently verified until recently.

The Magellanic Clouds (MC) are a most unique example where a number of new satellites were found by deep imaging surveys (e.g., \citealt{Drlica-Wagner2015,Martin2015,Nidever2017,Koposov2018,Torrealba2018,Cerny23}), providing a great opportunity to test $\Lambda$CDM models. Still, a deeper insight into the satellite population around nearby dwarfs is needed to put stronger constraints on cosmological models at small scales and understand how interaction and accretion events impact the evolution of dwarf galaxies, affect their morphology and kinematics, and possibly provide a viable mechanism to trigger the inflow of gas and the onset of starbursts \citep{Bekki2008,Stierwalt2015,Carlin2016,Kado-Fong2020}.

\begin{table*}[]
\caption{Main properties of the analyzed galaxies. Distances, $B$-magnitudes, and extinctions are from \cite{Annibali2020}. References for stellar and gas masses are: \cite{Cignoni2019} for NGC~5238, \cite{McQuinn2010} and \cite{Lelli2014} for UGC~6456, \cite{Berg2012} and \cite{Zheng2024} for UGC~6541, UGC~7605, and UGC~8638, \cite{Weisz2011} for UGC~8760.
}
\begin{center}
\begin{tabular}{cccccccccc}
    \toprule
    \midrule
    \addlinespace[0.3em]
    \multirow{2}{*}{Galaxy} & \multirow{2}{*}{Other name} & \multirow{2}{*}{Classification} & $l$ & $b$ & Distance & $M_B$ & $A_V$ & M$_{\ast}$ & M$_{HI}$ \\
    \addlinespace[0.3em]
    & & & [deg] & [deg] & [Mpc] & [mag] & [mag] & [$10^7$ M$_{\odot}$] & [$10^7$ M$_{\odot}$] \\
    \midrule
    NGC 5238 & UGC 8565 & SAB(s)dm & 107.41 & 64.19 & $4.51\pm0.14$ & $-$14.3 & 0.027 & $14\pm6$ & $2.9\pm0.7$ \\
    UGC 6456 & VII Zw 403 & BCD & 127.84 & 37.33 & $4.30\pm0.10$ & $-$13.7 & 0.102 &  $6.8 \pm 2.0$ & $4.5\pm0.5$ \\
    UGC 6541 & Mrk 178 & Im & 151.90 & 63.28 & $3.89\pm0.64$ & $-$13.6 & 0.052 & $1.20\pm0.35$ & $1.1\pm0.3$ \\
    UGC 7605 & & Im & 151.00 & 80.14 & $4.43\pm0.57$ & $-$13.5 & 0.040 &  $1.32\pm0.73$ & $2.1\pm0.2$ \\
    UGC 8638 & & Im & 23.27 & 78.99 & $4.27\pm0.34$ & $-$13.1 & 0.036 & $3.72\pm0.53$ & $1.9\pm0.2$ \\
    UGC 8760 & DDO 183 & Im & 77.79 & 73.45 & $3.24\pm0.34$ & $-$13.1 & 0.045 & $4.69\pm1.10$ & $2.1\pm0.3$ \\
    \bottomrule
\end{tabular}
\end{center}
\label{tab:table}
\end{table*}

By definition, satellites of dwarfs (or their disrupted relics) should be very faint and low-surface brightness, hence very hard to detect.
Consequently, direct observational evidence for merging events in dwarfs beyond the Local Group is limited to a few examples, such as the Magellanic irregular NGC~4449 \citep{Martinez-Delgado2012}, the extremely metal-poor DDO~68 \citep{Annibali2016}, or the sample of interacting dwarfs by  \cite{Paudel2018}, all systems  with relatively large stellar masses (M$_{\star}\sim10^8-10^9$ M$_{\odot}$). Moreover, the Local Group is quite an evolved system, and dwarfs in younger, less evolved ones should be studied as well.

To make improvements in this direction, we carried out a systematic search of the traces of merger events around dwarf galaxies. The Smallest Scale of Hierarchy Survey (SSH; \citealt{Annibali2020}) was specifically designed to determine the frequency and properties of interaction and merging events around a large sample of dwarf galaxies, by exploiting the high sensitivity and large field of view ($\simeq 23\arcmin \times 23\arcmin$) of the Large Binocular Camera (LBC) on the Large Binocular Telescope (LBT). We acquired deep $g$ and $r$ images for 45 late-type dwarfs at distances between $\sim 1$ and $\sim 10$ Mpc with the aim of revealing the presence of faint tidal features around them down to a surface brightness of  $\mu_r \simeq 31$ mag/arcsec$^2$ (where $r$ is in the SDSS system, thus AB mag, as in the rest of the paper). For the nearest galaxies in the sample, within a distance of $\sim$5 Mpc, this can be efficiently accomplished by selecting in the color-magnitude diagrams (CMDs) individual stars associated with the dwarf galaxy or with a potential satellite, and then exploring their spatial distribution. 

In this paper, we present the most striking examples of tidal and accretion features identified in six dwarfs (NGC~5238, UGC~6456, UGC~6541, UGC~7605, UGC~8638, and UGC~8760) out of the entire SSH sample. These galaxies, with a stellar masses between $\sim10^7-10^8$ M$_{\odot}$, represent some of the smallest systems observed so far bearing signs of merging events with one or more smaller companions.  


\section{Tidal features in dwarfs galaxies} 
\label{sec:data}

Fig.~\ref{fig:nicegal} shows portions of the LBT images centered on the six galaxies with tidal features, whose properties are summarized in Table \ref{tab:table}.  
In the plot, gray-scale tones are used to highlight the galaxies' low surface brightness emission, while $g$, $r$ color-combined images illustrate the morphology of the central star forming regions.

The detected tidal features appear as low surface brightness asymmetrical structures protruding from the main body of the galaxies and extending up to $\sim$2-3 kpc away from the center. They are reminiscent of shells (UGC 7605), giant plumes (UGC 6541, NGC 5238 south, UGC 8638, UGC 6456) and ``umbrella''-like features (NGC 5238 north, UGC 8760), similar to those described in \citet{Martinez-Delgado2010} in their classification of the most common signatures of mergers of satellites occurring in real and simulated (large) galaxies. 

In UGC~6456, UGC~6541, and UGC~8638 the substructures appear aligned with the galaxy major axis, protruding from both opposite sides of the main body. This resembles the disc-like substructures observed by \citet{Bellazzini2011} in UGC~4879, but, in contrast to that case, here there is always a significant asymmetry, in shape, extension and/or surface density, between the substructures at the opposite sides of the same galaxy.
In UGC~8760, a ``clump'' of stars is detected to the northeast, which appears connected to the main galaxy by a sort of ``bridge'', while a less prominent over-density is visible in the southern part of the galaxy. A similar morphology is revealed in NGC~5238, with an umbrella feature to the north connected to the main galaxy by a bridge, and a more sparse overdensity of stars to the south. 
In UGC~7605, the shell-like feature seems to wrap the southern part of galaxy, outlining what appears as a sort of ``fan''. These kind of low surface brightness substructures are commonly interpreted as the result of a merging event with a smaller satellite  which becomes tidally disrupted while still in orbit around the main galaxy \citep{Bullock2005,Johnston2008,Martinez-Delgado2010,Hong2020,Vera2022,Pascale2024}

None of these features are detected in existing SDSS images, lacking sufficient depth, and revealing only the star-forming central regions, while available HST images
of the six dwarfs cover relatively small areas around the galaxies' center, thus lacking the spatial coverage necessary to capture the tidal features we see in the distant outskirts.

\subsection{Color-magnitude diagrams} \label{sec:cmd}

\begin{figure}
\centering
\includegraphics[width=\linewidth]{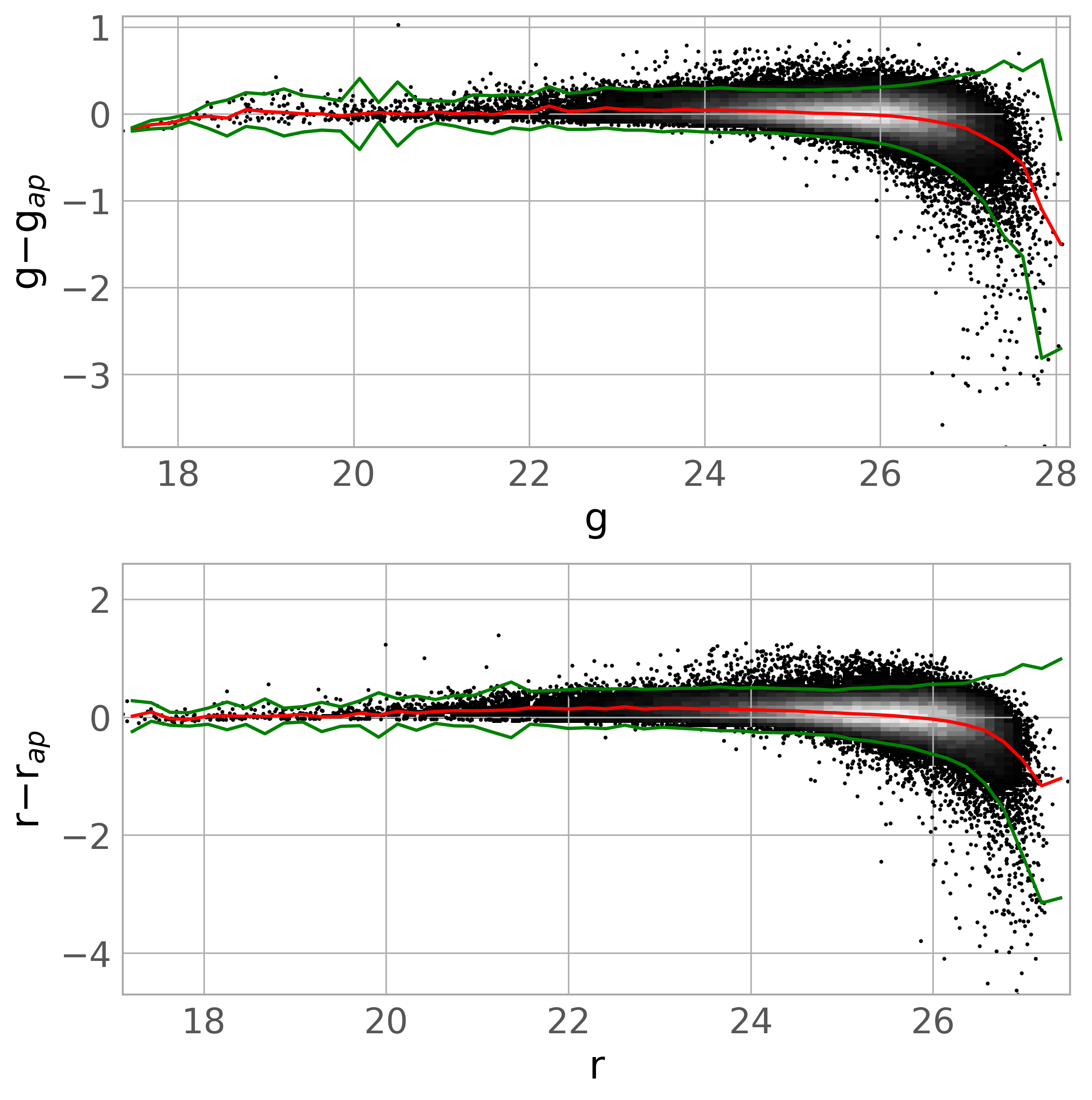}
\caption{One of the selection parameters we used on the photometric catalogue of UGC 8638, showing the PSF-fitting minus aperture magnitudes versus PSF magnitudes in $g$ (top panel) and $r$ (bottom panel). The red line shows the median of the distribution, while the green lines the $\pm 2\sigma$ levels, used to select the sources.}
\label{fig:cut}
\end{figure}

Thanks to the galaxies' relatively close distances of $\sim$3.2 - 4.5 Mpc, individual stars are resolved in our LBT images down to the brightest portion of the red giant branch (RGB, see the CMDs in Fig.~\ref{fig:cmds-all}).  
Photometric catalogues were obtained by performing Point Spread Function (PSF) fitting photometry with \textsc{PSFex} \citep{bertin13} on the stacked SSH $g$ and $r$ images, as described in detail in \citet[][]{Annibali2020}. The final catalogues contain calibrated astrometry and $g$,$r$ photometry\footnote{In the photometric system of the Sloan Digital Sky Survey \citep{Fukugita96}.}, plus several structural and quality parameters for all the detected sources. 
We applied various selection cuts to remove spurious detections and contaminants, using the available output parameters provided by \textsc{PSFex}. In particular we retained only sources with the following parameters:
\begin{itemize}
    \item flag = 0 in both bands, that is, best-measured individual sources;
    \item $g-g_{ap}$, $r-r_{ap}$ (i.e. PSF-fitting minus aperture magnitudes) within $2\sigma$ of the locus of point sources (see Figure \ref{fig:cut} for an example on UGC~8638; for some of the galaxies with less statistics, we used $3\sigma$ instead);
    \item half-light radius ($r_h$ in both $g$ and $r$) within $3\sigma$ of the mean value for point-sources.
\end{itemize}

\noindent
The last two selection criteria are aimed at removing background galaxies from our catalogues, as they are the main source of contamination in the range of magnitudes considered in our study (see \citealt{Bellazzini2011} and references therein). These final cleaned catalogues were then used for our selection of stellar counts in the six dwarfs, as described in the following.

In Fig.~\ref{fig:cmds-all} we show the CMDs derived for portions of the LBT images enclosing the main body of the six galaxies, together with PARSEC isochrones \citep{Bressan2012} from 10 Myr to 10 Gyr overlapped as an example on top of UGC 6541, and properly shifted according to the galaxy's  distance and foreground extinction (see Table \ref{tab:table}); the isochrones’ metallicity of Z = 0.001 is consistent with the H II region oxygen abundance of 12 + log(O/H) = 7.82 \citep{Berg2012}. 
All the CMDs are quite similar as all galaxies have similar distance and are of similar morphological type. The younger isochrones are bluer than the bulk of the observed stars because the contribution of young stars to the blue side of the CMDs is, at most, marginal. This is due to the circumstance that at distances of $\sim$3.3 - 4.3 Mpc the galaxies' innermost regions are unresolved in our LBT images because of stellar crowding. Thus the younger population, prominent but largely confined in the most crowded regions, is poorly  represented in our catalogues.
The handful of stars brighter than $r\simeq 23.0$ or redder than $g-r \simeq 1.5$ are mostly foreground stars of our own Galaxy or red compact background galaxies. Instead, the wedge-shaped distribution of sources with $g - r \la 0.8$, centered at $g - r < 0.5$ is dominated by completely unresolved background galaxies belonging to the blue sequence (see \citealt{{Bellazzini2011}} for discussion and references, in particular their Fig.~5).
On the other hand, stars fainter than $r\simeq 24.0$ and redder than $g - r \simeq 0.8$ are mostly genuine RGB stars of the SSH galaxies, with some asymptotic giant branch (AGB) stars beyond the evident discontinuity in  the luminosity function marking the RGB tip (between $r\simeq 24.0$ and $r\simeq 25.0$, depending on the specific galaxy). The isochrones are redder than the actual RGB stars because the old population is likely more metal poor than the HII regions where the metallicity was derived.

\begin{figure}
\centering
\includegraphics[width=0.8\linewidth]{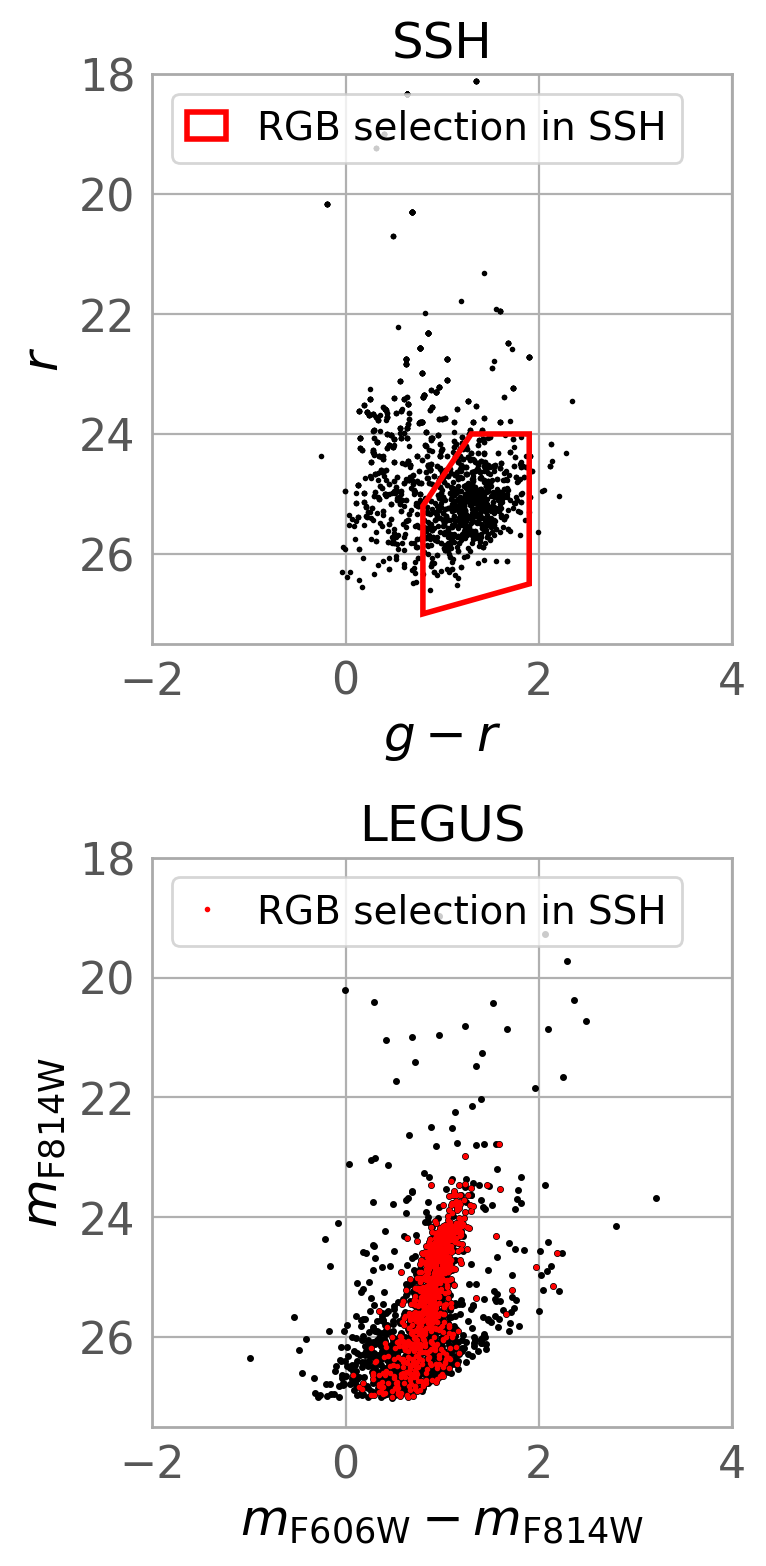}
\caption{CMD of NGC ~5238 after the match between SSH and LEGUS catalogs. \textit{Top Panel.} $r$ vs $g-r$ CMD (from SSH data); the red box shows the selection we used to isolate RGB stars. \textit{Bottom Panel.}  $m_{\mathrm{F814W}}$ vs $m_{\mathrm{F606W}}-m_{\mathrm{F814W}}$ CMD of the same matched stars shown in the top panel, but measured with $HST$ by LEGUS; the red points are the stars falling into our RGB selection box.}
\label{fig:cmd}
\end{figure}

Fig.~\ref{fig:cmd}, top panel, shows the polygonal region adopted to select RGB stars (in NGC 5238, but similar regions were defined for the other galaxies), which were used to trace the spatial distribution of the population older than $\sim$1-2 Gyr in the galaxies' main body and outskirts.  
To verify that stars selected in this way are indeed bona fide RGBs, we cross-matched our SSH catalogues with $HST$ photometry that is publicly available for all these galaxies. An example of the outcome of these experiments is displayed in Fig.~\ref{fig:cmd}, where the CMDs of the stars in common between our catalogue of NGC~5238 and that based on $HST$ data from the LEGUS survey \citep{Calzetti2015} are compared\footnote{Notice that the full LEGUS CMD of the region (see e.g. \citealt{Cignoni2019}) spans a much larger range of magnitudes, both brighter and fainter than those measurable by LBT, as well as a prominent blue plume of young stars}. In the LEGUS CMD, the stars that we selected as RGBs based on the SSH CMD are plotted in red and are fully confirmed as genuine RGB stars. Note that all the available $HST$ observations sample the central regions of these galaxies, hence the overlap with the SSH sample typically occurs in the innermost corona that is resolved in the SSH images; because of that, the photometric errors for the displayed CMDs are the largest, while both the photometric accuracy and the completeness are expected to improve in the outermost galaxy regions, where merger signatures are detected. 

\begin{figure*}
\centering
\includegraphics[width=\linewidth]{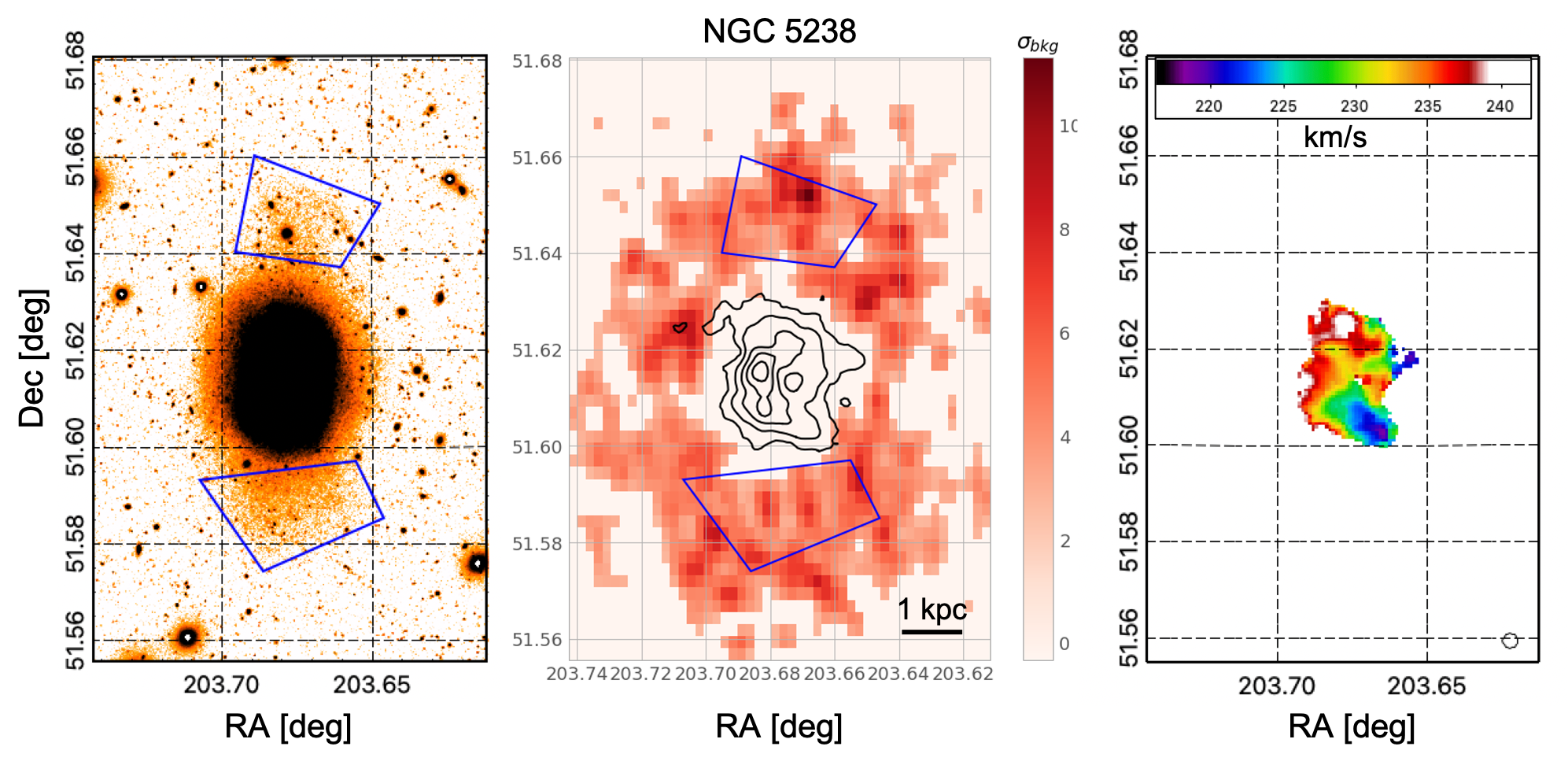}
\caption{\textit{Left Panel.} Central part of the $g$-band LBT image of NGC 5238, covering a field of view of $\sim 4.9\arcmin \times 7.5\arcmin$, or $\sim 6.0 \times 9.2$~kpc$^2$ (using the distance in Table \ref{tab:table}); the low surface brightness peculiar structures are highlighted by the blue polygons. \textit{Middle Panel.} Map of RGB stars in the same area, background-subtracted and with pixels below $3\sigma$ of the background set to zero; the HI iso-density contours are superimposed in black, and correspond to a column density of [2.5, 5, 10, 20] $\times 10^{20}$ atoms/cm$^2$. The black horizontal bar is 1 kpc long, adopting the distances in Table \ref{tab:table}. \textit{Right Panel.} HI velocity field; the beam size is also shown in the bottom-right corner. North is up, East is left. HI data are from the VLA program 23A-195 (Cannon, Schisgal et al. in preparation). All three panels are exactly on the same scale.}
\label{fig:ngc5238}
\end{figure*}

\begin{figure*}
\centering
\includegraphics[width=\linewidth]{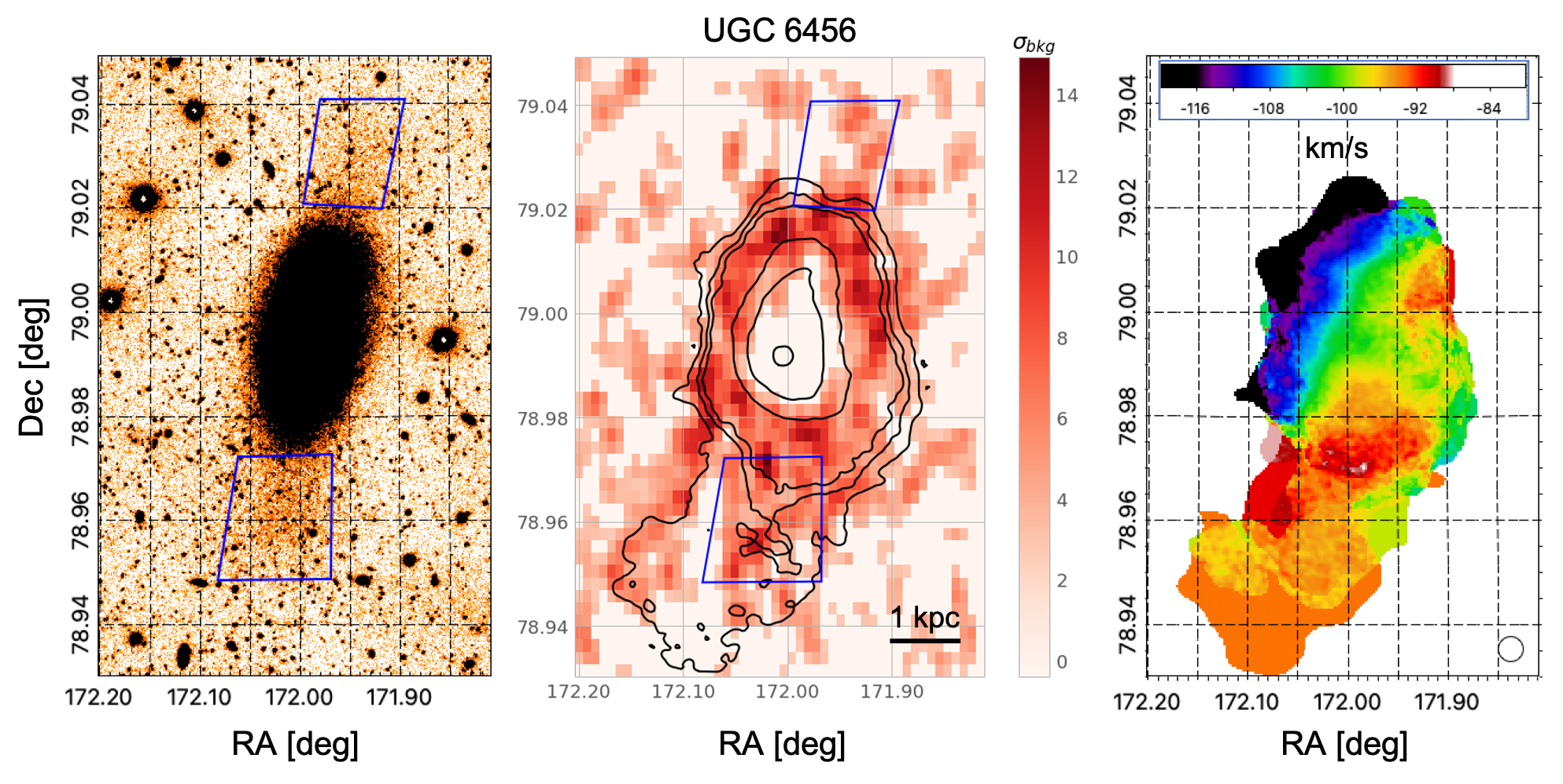}
\caption{Same as Figure \ref{fig:ngc5238} for UGC 6456.The field of view here is $\sim 4.5\arcmin \times 7.1\arcmin$, or $\sim 5.6 \times 8.9$~kpc$^2$, and the HI contours correspond to [0.1, 0.5, 1, 5.2, 10, 20] $\times 10^{20}$ atoms/cm$^2$. HI data are from the VLA program LITTLE THINGS \citep{Hunter2012}.}
\label{fig:ugc6456}
\end{figure*}

\begin{figure*}
\centering
\includegraphics[width=\linewidth]{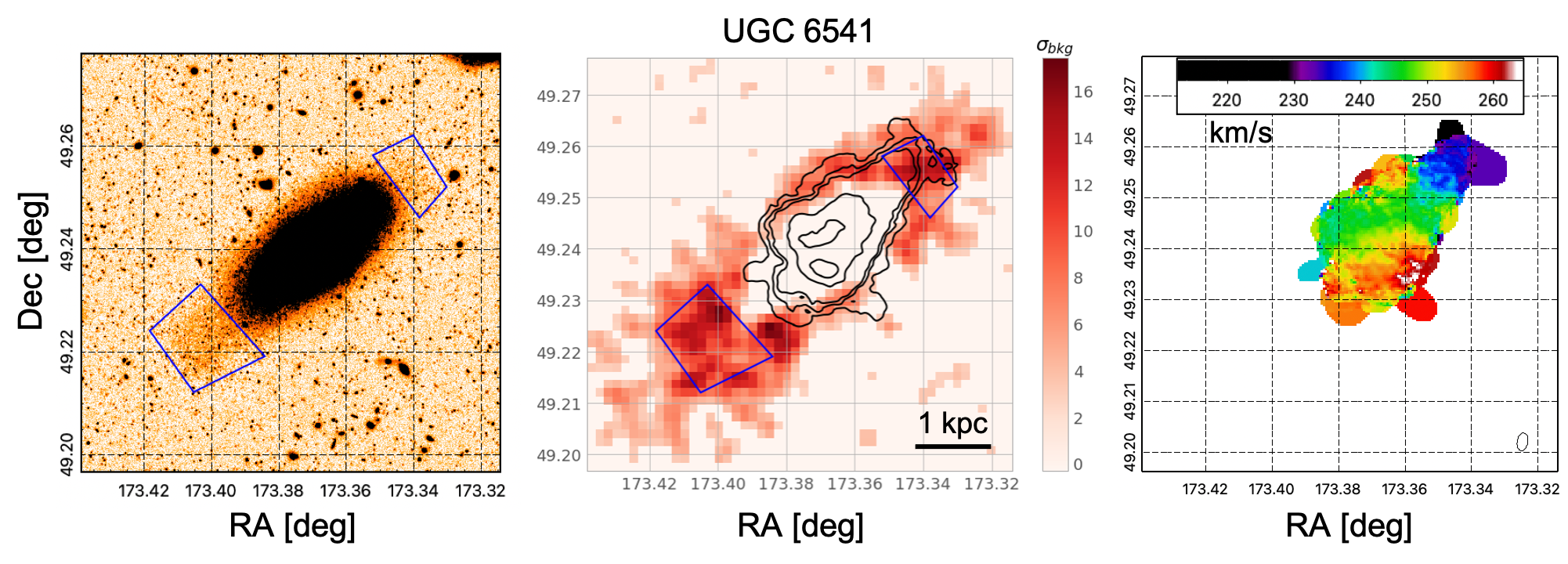}
\caption{Same as Figures \ref{fig:ngc5238} and \ref{fig:ugc6456} for UGC 6541. The field of view here is $\sim 4.9\arcmin \times 4.9\arcmin$, or $\sim 5.5\times 5.5$~kpc$^2$, and the HI contours correspond to [0.1,~0.5,~1,~5,~10]~$\times 10^{20}$ atoms/cm$^2$. HI data are from the VLA program LITTLE THINGS \citep{Hunter2012}.}
\label{fig:ugc6541}
\end{figure*}

\begin{figure*}
\centering
\includegraphics[width=\linewidth]{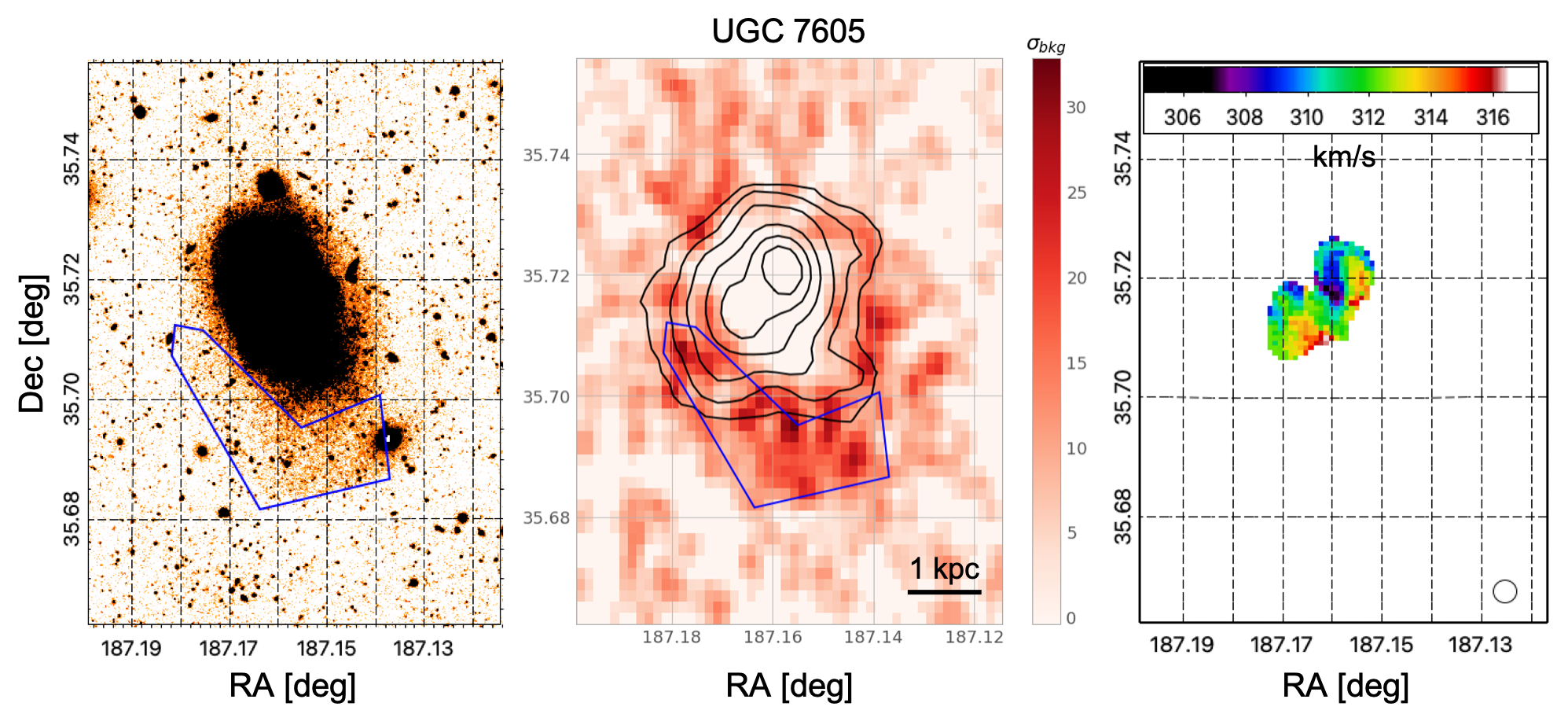}
\caption{Same as Figures \ref{fig:ngc5238} to \ref{fig:ugc6541} for UGC 7605. The field of view here is $\sim 4.1\arcmin \times 5.8\arcmin$, or $\sim 5.3 \times 7.4$~kpc$^2$, and the HI contours correspond to [0.1,~0.5,~1,~2,~3,~4]~$\times 10^{20}$ atoms/cm$^2$. Notice that the gas emission in the central and right panels have different resolution, with a $43\arcsec \times 38\arcsec$ beam for the HI contours, and a higher-resolution $16\arcsec \times 12\arcsec$ beam for the velocity field. HI data are from the GMRT program FIGGS \citep{Begum2008}.}
\label{fig:ugc7605}
\end{figure*}

\begin{figure*}
\centering
\includegraphics[width=\linewidth]{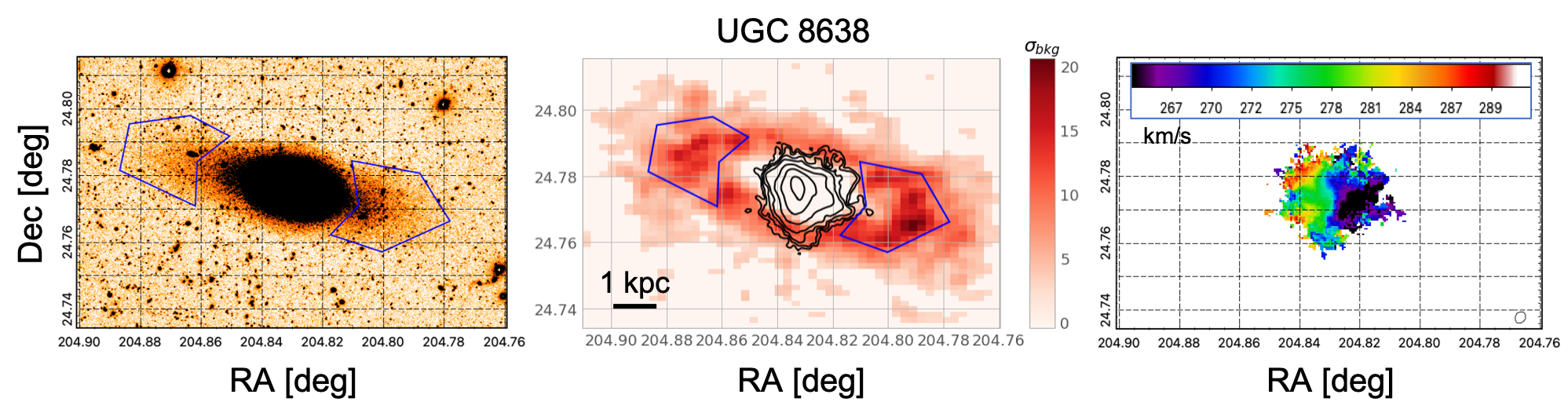}
\caption{Same as Figures \ref{fig:ngc5238} to \ref{fig:ugc7605} for UGC 8638. The field of view here is $\sim 8.3\arcmin \times 4.9\arcmin$, or $\sim 10.3 \times 6.1$~kpc$^2$, and the HI contours correspond to [0.1,~0.2,~1,~2,~5,~10,~20]~$\times 10^{20}$ atoms/cm$^2$. HI data are from the VLA \citep{Hunter2023}.}
\label{fig:ugc8638}
\end{figure*}

\begin{figure*}
\centering
\includegraphics[width=\linewidth]{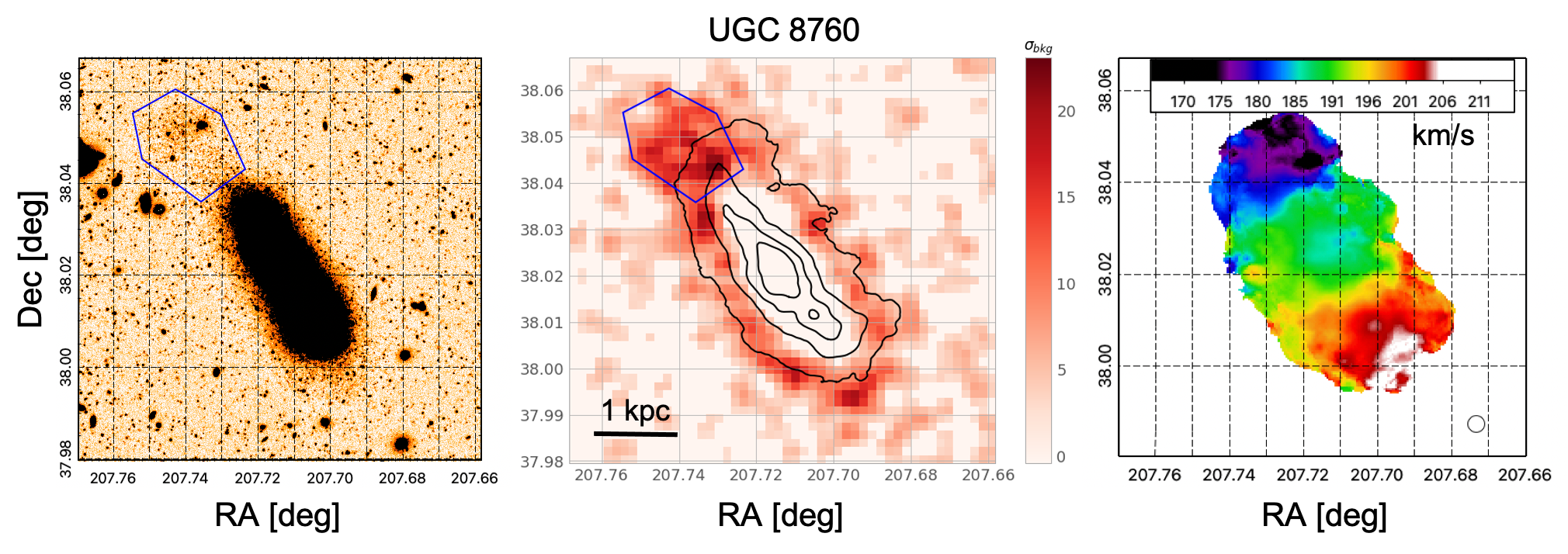}
\caption{Same as Figures \ref{fig:ngc5238} to \ref{fig:ugc8638} for UGC 8760. The field of view here is $\sim 5.3\arcmin \times 5.3\arcmin$, or $\sim 4.9 \times 5.0$~kpc$^2$, and the HI contours correspond to [1,~5,~10,~15]~$\times 10^{20}$ atoms/cm$^2$. HI data are from the VLA program ANGST \citep{Ott2012}.}
\label{fig:ugc8760}
\end{figure*}

\subsection{RGB star maps  and comparison with HI} \label{sec:RGBs}

In Figures~\ref{fig:ngc5238} to \ref{fig:ugc8760}, middle columns, we present the background-subtracted RGB density maps derived for the six dwarfs through the approach discussed in section~\ref{sec:cmd}. 
We estimated the background in an external portion of the total LBT field far from the main galaxy, computing the density of sources falling within our RGB selection polygon; to do so, we selected a region with the same area as the one centered on the galaxy, but located at least 10 arcmin far away from it, then selected candidate RGB stars in the CMDs of both fields, and computed the average surface density of the galaxy ($\mu_{gal}$), that of the background ($\mu_{bkg}$) and its sigma ($\sigma_{bkg}$). The RGB map is then created in units of $(\mu_{gal}-\mu_{bkg})/\sigma_{bkg}$.

Moreover, pixels below three times the standard deviation of the background ($\sigma_{bkg}$) were  set to zero in order to mask density noise due to background contaminants. Note that the ``hole'' in the central regions of each density map is an artefact due to the extreme crowding from star forming regions, as already mentioned above. The displayed surface density is in units of $\sigma_{bkg}$.

In the same figures, the RGB maps are compared with portions of our LBT images on the same scale (left columns), as well as with HI emission contours (middle columns) and HI velocity fields (right columns) obtained with the Karl G. Jansky Very Large Array (VLA) and with the Giant Metrewave Radio Telescope (GMRT) in the context of different programs or public surveys:  the Local Irregulars That Trace Luminosity Extremes in
The H I Nearby Galaxy Survey (LITTLE THINGS, \citealt{Hunter2012}) for UGC~6456 and UGC~6541;
the VLA Survey of ACS Nearby Galaxy Survey Treasury Galaxies (VLA-ANGST, \citealt{Ott2012}) for UGC~8760; the Faint Irregular Galaxies GMRT Survey (FIGGS, \citealt{Begum2008}) for UGC~7605; the VLA data of UGC~8638 are from \cite{Hunter2023}, while the VLA B configuration data of NGC~5238 
\footnote{VLA configuration C data of NGC~5238 were presented in \cite{Cannon2016}.} 
are from program VLA/23A-195 (P.I. Cannon), and will appear in a forthcoming publication (Cannon, Schisgal et al. in preparation).

Figures~\ref{fig:ngc5238} to \ref{fig:ugc8760} show that most of the external tidal features identified in the LBT images correspond to overdensities in the RGB maps; this is very clear for UGC 6541, UGC 7605, UGC 8638, and UGC 8760 (Figures \ref{fig:ugc6541} to \ref{fig:ugc8760}), while less evident in NGC 5238 (Fig. \ref{fig:ngc5238}) and UGC 6456 (Fig. \ref{fig:ugc6456}), whose maps are more noisy and with less striking overdensities. However, the substructures appear more extended in the RGB maps than in the $g$ (or $r$) images, a consequence of the fainter surface brightness probed by individual star counts compared to mere inspection of the images. On the other hand, none of the extended stellar features in the six dwarfs  appear to have a counterpart in HI (the only possible exception being the southern plume in NGC~6456), although deeper HI observations could potentially reveal low density gas emission at those locations.

In NGC~5238 (Fig.~\ref{fig:ngc5238}), the RGB map traces an extended distribution of old stars out to a galactocentric distance of $\sim$4 kpc. The most prominent over-density in the maps (more than 10 times the background standard deviation) corresponds to the northern shell visible in the LBT image.
As already discussed in \cite{Cannon2016}, the HI disc of NGC~5238 is asymmetric in the outer regions, with the morphological  major axis elongated from northeast to southwest. The high column density HI gas displays a crescent-shaped morphology. The galaxy also shows complex kinematics, with significant velocity asymmetries throughout the disk and an S-shaped velocity profile. The HI emission is very compact compared to the extended distribution of RGB stars, and there is not clear evidence for extended HI gas in the outer regions (also confirmed by new D configuration data currently under analysis by Cannon, Schisgal et al. in preparation). We further note that within the northern shell, there are at least three round extended sources that appear as good candidate star clusters (see the upper blue polygon in Fig.~\ref{fig:ngc5238}) that may be associated to the tidal feature. This intriguing hypothesis requires a more detailed analysis that we defer to a future contribution.

In UGC~6456 (Fig.~\ref{fig:ugc6456}), the extended southern and northern stellar plumes identified in the LBT images correspond to peaks in the RGB counts, although here the maps are quite noisy, also due to a significant incompleteness and contamination from foreground stars. 
The HI contours and velocity field  clearly show a highly distorted gas, in particular on the southern side, where the RGB density map exhibits a corresponding plume of stars. On the other hand, the northern stellar feature appears devoid of HI gas.  Because of its peculiar HI properties, UGC 6456 was already suggested to have undergone a possible interaction event, although no obvious HI or stellar companion was ever detected \citep{Ashley2017}. 

For UGC~6541 (Fig.~\ref{fig:ugc6541}), the least massive galaxy presented here, RGB stars trace very effectively the two tidal features oriented along its major axis. These appear significantly more extended in the RGB maps than in the LBT images, the one to the south-east reaching out to $\sim$3.5 kpc from the galaxy center.  
The HI gas extends toward the less prominent stellar plume to the northwest of the galaxy, while it is absent in the denser one to the southeast.  
The HI velocity field suggests complex large-scale dynamics. There is a modest velocity gradient ($\sim$10 km/s) within the inner disk that is co-spatial with the high surface brightness stellar component, although the central isovelocity contour ($\sim$ 250 km/s) is curved and is not aligned with either the optical major or the optical minor axes. The projected velocity gradient turns over (to higher velocities) in both the northern and the southern portions of the disk. Further, a gas component at lower velocity ($\sim$230 km/s) is present on the northwest side of the disk (co-spatial with, but larger than, the low surface brightness stellar extension in the same direction). Taken together, the complex HI kinematics suggest that UGC~6451 is involved in an ongoing merger event in which the neutral gas has not yet settled, as previously suggested by \cite{Ashley2017}.

In Figure \ref{fig:ugc7605}, the RGB map of UGC~7605 highlights a concentration of stars in correspondence of the southern fan-like feature detected in the LBT image. This is the most prominent  overdensity identified in the RGB map of UGC~7605, reaching more than 30 times the background standard deviation. 
Also in this case, the HI contours, derived from low-resolution data cubes ($43\arcsec \times 38\arcsec$ beam) of  the FIGGS data, indicate absence of gas at the location of the southern feature. The velocity field to the right, displayed for a higher resolution configuration ($16\arcsec \times 12\arcsec$ beam) of the FIGGS data, appears very complex and is highly distorted. 

In UGC~8638 (Figure~\ref{fig:ugc8638}), the two stellar plumes that extend to the northeast and southwest of its main central body are very clearly seen  both in the $g$-band image and in the RGB map. Here the western feature corresponds to an over-density exceeding 20 times the background standard deviation. 
Both features appear more extended in the RGB map than in the images, reaching out to $\sim$4 kpc from the galaxy center.  
The HI emission map and velocity field come from VLA data presented in \cite{Hunter2023}. 
The gas is confined to the galaxy main body, and the velocity field exhibits significant distortion. 

Lastly, we present in Figure \ref{fig:ugc8760} the RGB map of UGC~8760. 
This dwarf irregular, the closest in our sample, exhibits a prominent RGB over-density at the northeast of its central body, corresponding to the faint clump of stars visible in the LBT image. The RGB map clearly shows that this clump of stars is not detached from the galaxy, but rather represents the densest part of an elongated, continuous distribution departing from the galaxy itself. 
Another smaller over-density is visible to the south, where an even fainter arc of stars appears at the edge of the HI emission.
The velocity field is quite regular except for a distortion on the northern side probably due to the interaction with the smaller satellite that is responsible  for the northern over-density, in the area included in the blue polygon plotted in the left panel (RA$\simeq 207.74$, Dec$\simeq 38.05$); the satellite is not completely merged yet, but there is a clear bridge of stars connecting it to the main body of the galaxy, corresponding to the highest peak in the RGB map.

We notice that although the tidal features around the six dwarfs are cleanly detected, and in some cases even more clearly seen directly in the LBT images, our RGB map analysis demonstrates that they are made of old stars (see also Appendix~\ref{app:stars}): this proves that {\it i)} they are statistically significant in terms of overdensities in star counts with respect to the background, and {\it ii)} they are not vagaries of recent star formation episodes but genuine results of the perturbation of the gravitational potential.



\section{Discussion}

The LBT images of the six dwarfs analyzed in this paper unequivocally reveal asymmetric low surface density structures emanating from the main body of the galaxies, with characteristic scales of $\sim$2-3~kpc or more, and shapes consistent with those predicted by simulations for relics of merger events \citep[][see also \citealt{Pascale2024} for detailed hydrodynamical N-body simulations of UGC~8760 and NGC~5238]{{Bullock2005,Johnston2008,Martinez-Delgado2010,Hong2020,Vera2022}}. 
These substructures are largely dominated by RGB stars, whose old ages ($\ga 2$~Gyr) guarantee that they are not due to recent off-center star formation episodes but instead are due to genuine perturbations of the gravitational potential. Indeed, the analysis of the RGB spatial distribution reveals that these substructures are even more extended than what is directly visible in the LBT images.

For all our six galaxies, not only the RGB star distribution, but also the gas morphology and kinematics hold the signature of merging/interaction events. 
The HI intensity maps typically show an asymmetric and distorted morphology, and the HI velocity fields are highly disturbed. For NGC~5238, UGC~6456, and UGC~6541, these peculiar HI properties were already pointed out and discussed by previous studies \citep{Cannon2016, Ashley2017}, which proposed a merger event as a viable explanation  but could not find any distinct companion obviously interacting with the dwarfs.

Since the gas has a relatively short-term memory of past mergers, the high level of perturbation observed in HI provides a stringent constraint on how recently the merger/interaction event may have occurred. On a first approximation, the interaction has likely occurred no longer than $\sim$1 Gyr ago since, on longer timescales, orbital mixing and energy dissipation tends to erase high-signal distortions in the HI velocity field. Indeed, hydro-dynamical N-body simulations for UGC~8760 and NGC~5238 \citep{Pascale2024} show that the low surface brightness substructures observed in these galaxies, as well as the gas properties, are well explained with recent ($\lesssim$ 1 Gyr ago) interactions with smaller companions with stellar mass a few 10$^5$ M$_{\sun}$, $\sim$50 times less massive than the corresponding hosts. Note that all six dwarfs are located 1-2 Mpc away from large galaxies, therefore we can safely exclude that the observed stellar tidal features and HI peculiarities are due to the interaction with a nearby massive companion. 
 
We expect the occurrence of a recent interaction event to have left its signature also on the dwarfs' star formation history (SFH). In fact, hydro-dynamical simulations predict that the merger of a gas-rich dwarf with a smaller companion will cause bursts of star formation at, or just after, pericenter passages \citep{Bekki2008,Starkenburg2016a,Starkenburg2016b}. 
For NGC~5238, \cite{Cignoni2019} derived a clear increase in the star formation activity around 20 Myr ago,  while the SFH is more uncertain at earlier epochs. The SFH of UGC~6456, as inferred from optical CMDs by \cite{McQuinn2010}, shows a very recent burst in the last 100 Myr, and 3 lower peaks over the last 1 Gyr. UGC~6541 is an actively star forming galaxy, with a prominent population of young, Wolf-Rayet stars \citep{Kehrig2013} and evidence, from resolved CMD analysis, of vigorous star formation occurred several hundreds Myr ago \citep{Schulte2000}. The SFH of UGC~8760 derived by \cite{Weisz2011} exhibits a modest peak of star formation activity about 2  Gyr ago, although the errors are quite large. These SFHs provide a precious ingredient to be combined with hydrodynamical N-body simulations in order to reconstruct a coherent picture for the dwarfs' merging history and its effect on the galaxies' star formation activity. 

It is very interesting to notice that out of the 45 dwarfs in the full SSH sample, the 6 presented here are those showing the clearest signs of recent interaction/merger events, i.e. a fraction of about F=13\%\footnote{This should be considered as a preliminary lower limit as there are a couple of cases that requires additional investigation. In any case the fraction of galaxies in the SSH sample showing clear signs of past merging event should not exceed 20\%.}. We can compare this rate to what is found in the literature. \cite{Kado-Fong2020} present an empirical study on unresolved galaxies using imaging from the Hyper Suprime-Cam Subaru Strategic Program and spectra from the Galaxy and Mass Assembly and Sloan Digital Sky Surveys campaigns. They build a sample of 6875 isolated galaxies with $\log(M_{\star}/M_{\odot})<9.6$ and use an automatic detection method to look for tidal features. They identify 226 galaxies with detectable low surface brightness debris (of which 101 with unambiguous signs of a dwarf–dwarf merger after visual inspection) making the detection rate about F=3.3\% (6.1\% when considering only galaxies at $z<0.05$). F strongly grows as a function of the star formation activity, i.e. for galaxies with bluer colors and extremely high H$\alpha$ equivalent widths, reaching about 15\% in starbursts.

Most recently, \cite{Lazar2024} analyzed a sample of 257 local dwarf galaxies living in low-density environments, finding that 14\%, 27\%, and 19\% of early-type, late-type and featureless dwarfs respectively, exhibit evidence of interactions; they also show how morphological parameters could be used to separate interacting from non-interacting dwarfs, in particular when taking into account the level of asymmetry. 
This is also evident in the analysis by \cite{Paudel2018}, who presented a catalog of 177 interacting dwarf galaxies with stellar masses $< 10^{10} M_{\odot}$ and redshift $< 0.02$; they clearly show and classify low surface brightness features around these galaxies, including shells, stellar streams, loops, and antennae, likely resulting from an interaction between dwarfs (more than 75\% of the galaxies reside in low density environments).

On the theoretical side, \cite{Deason2014} use a suite of cosmological zoom-in simulations of Milky Way- and M31-like host halos to study major mergers (stellar mass ratio $>= 0.1$, i.e. total mass ratio about $>0.3$); they find that the fraction of mergers is 
$\sim$ 3\% for $M_{\star}=10^{3-5}\ M_{\odot}$, $\sim$ 5\% for $M_{\star}=10^{5-6}\ M_{\odot}$, and 10\% for $M_{\star}>10^6\ M_{\odot}$/\. 
More isolated galaxies with $M_{\star}=10^{7-9}\ M_{\odot}$ which are within $z=0.5$ have F=8.9\% (within $z=1$, F=18.4\%).
However, we notice that these results apply to major mergers with mass ratios larger than 0.1, whereas the tidal features detected in the outskirts of our six dwarfs are likely due to minor mergers, as also supported by the simulations presented by \cite{Pascale2024} for UGC~8760 and NGC~5238, where the satellites are 50 times less massive than their  hosts.

Another interesting work by \cite{Martin2021} analyzed 1000 dwarf galaxies from the NEWHORIZON cosmological simulation, to investigate how mergers and fly-bys drive their mass assembly and structural evolution. Among the galaxies exhibiting disturbed morphologies, only a small proportion have disturbances driven by mergers; these features are instead primarily the result of interactions that do not end in a merger (e.g. fly-bys). In their sample, galaxies of all stellar masses spend significant amounts of time (between 10\% and 30\%) in a morphologically disturbed state, and the time that merger remnants take to relax increases towards low redshift. Though this study is focused on redshifts between 5 and 0.5, these results are highlighting the need for a local sample to detect and study disturbed galaxies.

More recently, \cite{Deason2022} modeled the merger history of dwarf galaxies with M$_{halo}\sim10^{10}\ M_{\odot}$ (for both major and minor mergers); given the consistent uncertainties on the stellar mass - halo mass ratio, our galaxies could fall in this regime. Their model B, characterised by a lower mass threshold for the ignition of star formation\footnote{$M_{50}=10^{7.5}~M_{\sun}$, with respect to $M_{50}=10^{9.3}~M_{\sun}$ adopted for the other set of models they explore (model A); see \cite{Deason2022} for definition and details.} within a $\Lambda$CDM framework, predicts a fraction of dwarfs with disturbed morphologies due to interaction with satellites of $\simeq 10\%$, in agreement with our findings.

We can conclude that three independent observational studies, targeting very different samples and using different methods, broadly agree that the fraction F of late type isolated dwarfs showing detectable signs of interaction/merging with satellites is between $\simeq$10\% and $\simeq$20\%, in good agreement with the theoretical predictions within the $\Lambda$CDM framework by \citet{Deason2014,Deason2022} and \cite{Martin2021}.
In this context, SSH provides the most ``local'' estimate of F. The nearby target galaxies offer the opportunity of a detailed simultaneous study of the morphology and the kinematics of the stars and the gas components, as well as of the star formation history, as derived from the CMD of resolved stars. This, in turn, allows detailed hydrodynamical modeling of the merging events that lead to the present status of the systems, giving an unmatched insight into the assembly process of dwarf galaxies \citep{Pascale2021,Pascale2022}. The detectability fraction discussed here of course depends on the depth of our observations, making new coming facilities, like the Roman telescope, really precious for this kind of analysis.


\section{Conclusions}

In this paper, we presented the six most convincing cases of tidal features discovered in our SSH survey aimed at studying the process of hierarchical merging at the smallest galactic scales \citep[SSH,][]{Annibali2020}. The six galaxies are isolated dIrrs or BCDs with stellar mass in the $10^7~M_{\sun}<M_{\star}<2\times 10^8~M_{\sun}$ range. 

The asymmetries and substructures we found in the distribution of old (RGB) stars extend beyond the main body of the target galaxies, out to a few kpc from the galaxies' centers, and are similar to the tidal features observed and classified by \cite{Martinez-Delgado2010} in giant spirals. When traced with RGB star counts all of them are found to have high statistical significance, as overdensities above a background of unrelated fore/background sources.
These features witness the effect of dynamical perturbations, and in these dwarfs that are distant from large (massive) companions they indicate an interaction with former satellites, marking an important milestone towards a deeper understanding of hierarchical merging phenomena at the smallest galaxy scales.



From the identification of six galaxies with highly convincing merger/interaction features out of the entire sample of 45 dwarfs in SSH, we can set our merger detection fraction at about F=13\%, in good agreement with recent independent studies and theoretical predictions \citep{Deason2014,Deason2022,Kado-Fong2020,Martin2021,Lazar2024}. 
However, a few, less obvious, other cases could be present in the remaining SSH sample, and their stellar and gaseous properties are currently under study, making our fraction a lower limit.



\begin{acknowledgements}
      The National Radio Astronomy Observatory is a facility of the National Science Foundation operated under cooperative agreement by Associated Universities, Inc. (AUI). E.S. wishes to thank Valentina La Torre and Maria Elisabetta Dalla Mura, invaluable sources of support and advice throughout the past years and during the work that brought to this publication.
\end{acknowledgements}


\bibliography{bib}

\begin{thebibliography}{57}
\expandafter\ifx\csname natexlab\endcsname\relax\def\natexlab#1{#1}\fi

\bibitem[{{Annibali} {et~al.}(2020){Annibali}, {Beccari}, {Bellazzini}, {Tosi}, {Cusano}, {Paris}, {Cignoni}, {Ciotti}, {Nipoti}, \& {Sacchi}}]{Annibali2020}
{Annibali}, F., {Beccari}, G., {Bellazzini}, M., {et~al.} 2020, \mnras, 491, 5101

\bibitem[{{Annibali} {et~al.}(2016){Annibali}, {Nipoti}, {Ciotti}, {Tosi}, {Aloisi}, {Bellazzini}, {Cignoni}, {Cusano}, {Paris}, \& {Sacchi}}]{Annibali2016}
{Annibali}, F., {Nipoti}, C., {Ciotti}, L., {et~al.} 2016, \apjl, 826, L27

\bibitem[{{Ashley} {et~al.}(2017){Ashley}, {Simpson}, {Elmegreen}, {Johnson}, \& {Pokhrel}}]{Ashley2017}
{Ashley}, T., {Simpson}, C.~E., {Elmegreen}, B.~G., {Johnson}, M., \& {Pokhrel}, N.~R. 2017, \aj, 153, 132

\bibitem[{{Begum} {et~al.}(2008){Begum}, {Chengalur}, {Karachentsev}, {Sharina}, \& {Kaisin}}]{Begum2008}
{Begum}, A., {Chengalur}, J.~N., {Karachentsev}, I.~D., {Sharina}, M.~E., \& {Kaisin}, S.~S. 2008, \mnras, 386, 1667

\bibitem[{{Bekki}(2008)}]{Bekki2008}
{Bekki}, K. 2008, \mnras, 388, L10

\bibitem[{{Bellazzini} {et~al.}(2011){Bellazzini}, {Beccari}, {Oosterloo}, {Galleti}, {Sollima}, {Correnti}, {Testa}, {Mayer}, {Cignoni}, {Fraternali}, \& {Gallozzi}}]{Bellazzini2011}
{Bellazzini}, M., {Beccari}, G., {Oosterloo}, T.~A., {et~al.} 2011, \aap, 527, A58

\bibitem[{{Berg} {et~al.}(2012){Berg}, {Skillman}, {Marble}, {van Zee}, {Engelbracht}, {Lee}, {Kennicutt}, {Calzetti}, {Dale}, \& {Johnson}}]{Berg2012}
{Berg}, D.~A., {Skillman}, E.~D., {Marble}, A.~R., {et~al.} 2012, \apj, 754, 98

\bibitem[{{Bertin}(2013)}]{bertin13}
{Bertin}, E. 2013, {PSFEx: Point Spread Function Extractor}, Astrophysics Source Code Library, record ascl:1301.001

\bibitem[{{Besla} {et~al.}(2018){Besla}, {Patton}, {Stierwalt}, {Rodriguez-Gomez}, {Patel}, {Kallivayalil}, {Johnson}, {Pearson}, {Privon}, \& {Putman}}]{Besla2018}
{Besla}, G., {Patton}, D.~R., {Stierwalt}, S., {et~al.} 2018, \mnras, 480, 3376

\bibitem[{{Blanton} \& {Moustakas}(2009)}]{Blanton2009}
{Blanton}, M.~R. \& {Moustakas}, J. 2009, \araa, 47, 159

\bibitem[{{Bressan} {et~al.}(2012){Bressan}, {Marigo}, {Girardi}, {Salasnich}, {Dal Cero}, {Rubele}, \& {Nanni}}]{Bressan2012}
{Bressan}, A., {Marigo}, P., {Girardi}, L., {et~al.} 2012, \mnras, 427, 127

\bibitem[{{Bullock} \& {Boylan-Kolchin}(2017)}]{Bullock2017}
{Bullock}, J.~S. \& {Boylan-Kolchin}, M. 2017, \araa, 55, 343

\bibitem[{{Bullock} \& {Johnston}(2005)}]{Bullock2005}
{Bullock}, J.~S. \& {Johnston}, K.~V. 2005, \apj, 635, 931

\bibitem[{{Calzetti} {et~al.}(2015){Calzetti}, {Lee}, {Sabbi}, {Adamo}, {Smith}, {Andrews}, {Ubeda}, {Bright}, {Thilker}, {Aloisi}, {Brown}, {Chandar}, {Christian}, {Cignoni}, {Clayton}, {da Silva}, {de Mink}, {Dobbs}, {Elmegreen}, {Elmegreen}, {Evans}, {Fumagalli}, {Gallagher}, {Gouliermis}, {Grebel}, {Herrero}, {Hunter}, {Johnson}, {Kennicutt}, {Kim}, {Krumholz}, {Lennon}, {Levay}, {Martin}, {Nair}, {Nota}, {{\"O}stlin}, {Pellerin}, {Prieto}, {Regan}, {Ryon}, {Schaerer}, {Schiminovich}, {Tosi}, {Van Dyk}, {Walterbos}, {Whitmore}, \& {Wofford}}]{Calzetti2015}
{Calzetti}, D., {Lee}, J.~C., {Sabbi}, E., {et~al.} 2015, \aj, 149, 51

\bibitem[{{Cannon} {et~al.}(2016){Cannon}, {McNichols}, {Teich}, {Ball}, {Banovetz}, {Brock}, {Eisner}, {Fitzgibbon}, {Miazzo}, {Nizami}, {Reilly}, {Ruvolo}, \& {Singer}}]{Cannon2016}
{Cannon}, J.~M., {McNichols}, A.~T., {Teich}, Y.~G., {et~al.} 2016, \aj, 152, 202

\bibitem[{{Carlin} {et~al.}(2016){Carlin}, {Sand}, {Price}, {Willman}, {Karunakaran}, {Spekkens}, {Bell}, {Brodie}, {Crnojevi{\'c}}, {Forbes}, {Hargis}, {Kirby}, {Lupton}, {Peter}, {Romanowsky}, \& {Strader}}]{Carlin2016}
{Carlin}, J.~L., {Sand}, D.~J., {Price}, P., {et~al.} 2016, \apjl, 828, L5

\bibitem[{{Cerny} {et~al.}(2023){Cerny}, {Drlica-Wagner}, {Li}, {Pace}, {Olsen}, {No{\"e}l}, {van der Marel}, {Carlin}, {Choi}, {Erkal}, {Geha}, {James}, {Mart{\'\i}nez-V{\'a}zquez}, {Massana}, {Medina}, {Miller}, {Mutlu-Pakdil}, {Nidever}, {Sakowska}, {Stringfellow}, {Carballo-Bello}, {Ferguson}, {Kuropatkin}, {Mau}, {Tollerud}, {Vivas}, \& {Delve Collaboration}}]{Cerny23}
{Cerny}, W., {Drlica-Wagner}, A., {Li}, T.~S., {et~al.} 2023, \apjl, 953, L21

\bibitem[{{Cignoni} {et~al.}(2019){Cignoni}, {Sacchi}, {Tosi}, {Aloisi}, {Cook}, {Calzetti}, {Lee}, {Sabbi}, {Thilker}, {Adamo}, {Dale}, {Elmegreen}, {Gallagher}, {Grebel}, {Johnson}, {Messa}, {Smith}, \& {Ubeda}}]{Cignoni2019}
{Cignoni}, M., {Sacchi}, E., {Tosi}, M., {et~al.} 2019, \apj, 887, 112

\bibitem[{{Deason} {et~al.}(2014){Deason}, {Wetzel}, \& {Garrison-Kimmel}}]{Deason2014}
{Deason}, A., {Wetzel}, A., \& {Garrison-Kimmel}, S. 2014, \apj, 794, 115

\bibitem[{{Deason} {et~al.}(2022){Deason}, {Bose}, {Fattahi}, {Amorisco}, {Hellwing}, \& {Frenk}}]{Deason2022}
{Deason}, A.~J., {Bose}, S., {Fattahi}, A., {et~al.} 2022, \mnras, 511, 4044

\bibitem[{{Diemand} {et~al.}(2008){Diemand}, {Kuhlen}, {Madau}, {Zemp}, {Moore}, {Potter}, \& {Stadel}}]{Diemand2008}
{Diemand}, J., {Kuhlen}, M., {Madau}, P., {et~al.} 2008, \nat, 454, 735

\bibitem[{{Dooley} {et~al.}(2017){Dooley}, {Peter}, {Carlin}, {Frebel}, {Bechtol}, \& {Willman}}]{Dooley2017}
{Dooley}, G.~A., {Peter}, A. H.~G., {Carlin}, J.~L., {et~al.} 2017, \mnras, 472, 1060

\bibitem[{{Drlica-Wagner} {et~al.}(2015){Drlica-Wagner}, {Bechtol}, {Rykoff}, {Luque}, {Queiroz}, {Mao}, {Wechsler}, {Simon}, {Santiago}, {Yanny}, {Balbinot}, {Dodelson}, {Fausti Neto}, {James}, {Li}, {Maia}, {Marshall}, {Pieres}, {Stringer}, {Walker}, {Abbott}, {Abdalla}, {Allam}, {Benoit-L{\'e}vy}, {Bernstein}, {Bertin}, {Brooks}, {Buckley-Geer}, {Burke}, {Carnero Rosell}, {Carrasco Kind}, {Carretero}, {Crocce}, {da Costa}, {Desai}, {Diehl}, {Dietrich}, {Doel}, {Eifler}, {Evrard}, {Finley}, {Flaugher}, {Fosalba}, {Frieman}, {Gaztanaga}, {Gerdes}, {Gruen}, {Gruendl}, {Gutierrez}, {Honscheid}, {Kuehn}, {Kuropatkin}, {Lahav}, {Martini}, {Miquel}, {Nord}, {Ogando}, {Plazas}, {Reil}, {Roodman}, {Sako}, {Sanchez}, {Scarpine}, {Schubnell}, {Sevilla-Noarbe}, {Smith}, {Soares-Santos}, {Sobreira}, {Suchyta}, {Swanson}, {Tarle}, {Tucker}, {Vikram}, {Wester}, {Zhang}, {Zuntz}, \& {DES Collaboration}}]{Drlica-Wagner2015}
{Drlica-Wagner}, A., {Bechtol}, K., {Rykoff}, E.~S., {et~al.} 2015, \apj, 813, 109

\bibitem[{{Fukugita} {et~al.}(1996){Fukugita}, {Ichikawa}, {Gunn}, {Doi}, {Shimasaku}, \& {Schneider}}]{Fukugita96}
{Fukugita}, M., {Ichikawa}, T., {Gunn}, J.~E., {et~al.} 1996, \aj, 111, 1748

\bibitem[{{Hunter} {et~al.}(2012){Hunter}, {Ficut-Vicas}, {Ashley}, {Brinks}, {Cigan}, {Elmegreen}, {Heesen}, {Herrmann}, {Johnson}, {Oh}, {Rupen}, {Schruba}, {Simpson}, {Walter}, {Westpfahl}, {Young}, \& {Zhang}}]{Hunter2012}
{Hunter}, D.~A., {Ficut-Vicas}, D., {Ashley}, T., {et~al.} 2012, \aj, 144, 134

\bibitem[{Hunter(2023)}]{Hunter2023}
Hunter, L.~C. 2023, PhD thesis

\bibitem[{{Jahn} {et~al.}(2019){Jahn}, {Sales}, {Wetzel}, {Boylan-Kolchin}, {Chan}, {El-Badry}, {Lazar}, \& {Bullock}}]{Jahn2019}
{Jahn}, E.~D., {Sales}, L.~V., {Wetzel}, A., {et~al.} 2019, \mnras, 489, 5348

\bibitem[{{Johnston} {et~al.}(2008){Johnston}, {Bullock}, {Sharma}, {Font}, {Robertson}, \& {Leitner}}]{Johnston2008}
{Johnston}, K.~V., {Bullock}, J.~S., {Sharma}, S., {et~al.} 2008, \apj, 689, 936

\bibitem[{{Kado-Fong} {et~al.}(2020){Kado-Fong}, {Greene}, {Greco}, {Beaton}, {Goulding}, {Johnson}, \& {Komiyama}}]{Kado-Fong2020}
{Kado-Fong}, E., {Greene}, J.~E., {Greco}, J.~P., {et~al.} 2020, \aj, 159, 103

\bibitem[{{Kehrig} {et~al.}(2013){Kehrig}, {P{\'e}rez-Montero}, {V{\'\i}lchez}, {Brinchmann}, {Kunth}, {Garc{\'\i}a-Benito}, {Crowther}, {Hern{\'a}ndez-Fern{\'a}ndez}, {Durret}, {Contini}, {Fern{\'a}ndez-Mart{\'\i}n}, \& {James}}]{Kehrig2013}
{Kehrig}, C., {P{\'e}rez-Montero}, E., {V{\'\i}lchez}, J.~M., {et~al.} 2013, \mnras, 432, 2731

\bibitem[{{Koposov} {et~al.}(2018){Koposov}, {Walker}, {Belokurov}, {Casey}, {Geringer-Sameth}, {Mackey}, {Da Costa}, {Erkal}, {Jethwa}, {Mateo}, {Olszewski}, \& {Bailey}}]{Koposov2018}
{Koposov}, S.~E., {Walker}, M.~G., {Belokurov}, V., {et~al.} 2018, \mnras, 479, 5343

\bibitem[{{Lazar} {et~al.}(2024){Lazar}, {Kaviraj}, {Watkins}, {Martin}, {Bichang'a}, \& {Jackson}}]{Lazar2024}
{Lazar}, I., {Kaviraj}, S., {Watkins}, A.~E., {et~al.} 2024, \mnras, 529, 499

\bibitem[{{Lelli} {et~al.}(2014){Lelli}, {Verheijen}, \& {Fraternali}}]{Lelli2014}
{Lelli}, F., {Verheijen}, M., \& {Fraternali}, F. 2014, \aap, 566, A71

\bibitem[{{Martin} {et~al.}(2021){Martin}, {Jackson}, {Kaviraj}, {Choi}, {Devriendt}, {Dubois}, {Kimm}, {Kraljic}, {Peirani}, {Pichon}, {Volonteri}, \& {Yi}}]{Martin2021}
{Martin}, G., {Jackson}, R.~A., {Kaviraj}, S., {et~al.} 2021, \mnras, 500, 4937

\bibitem[{{Martin} {et~al.}(2015){Martin}, {Nidever}, {Besla}, {Olsen}, {Walker}, {Vivas}, {Gruendl}, {Kaleida}, {Mu{\~n}oz}, {Blum}, {Saha}, {Conn}, {Bell}, {Chu}, {Cioni}, {de Boer}, {Gallart}, {Jin}, {Kunder}, {Majewski}, {Martinez-Delgado}, {Monachesi}, {Monelli}, {Monteagudo}, {No{\"e}l}, {Olszewski}, {Stringfellow}, {van der Marel}, \& {Zaritsky}}]{Martin2015}
{Martin}, N.~F., {Nidever}, D.~L., {Besla}, G., {et~al.} 2015, \apjl, 804, L5

\bibitem[{{Mart{\'\i}nez-Delgado} {et~al.}(2010){Mart{\'\i}nez-Delgado}, {Gabany}, {Crawford}, {Zibetti}, {Majewski}, {Rix}, {Fliri}, {Carballo-Bello}, {Bardalez-Gagliuffi}, {Pe{\~n}arrubia}, {Chonis}, {Madore}, {Trujillo}, {Schirmer}, \& {McDavid}}]{Martinez-Delgado2010}
{Mart{\'\i}nez-Delgado}, D., {Gabany}, R.~J., {Crawford}, K., {et~al.} 2010, \aj, 140, 962

\bibitem[{{Mart{\'{\i}}nez-Delgado} {et~al.}(2012){Mart{\'{\i}}nez-Delgado}, {Romanowsky}, {Gabany}, {Annibali}, {Arnold}, {Fliri}, {Zibetti}, {van der Marel}, {Rix}, {Chonis}, {Carballo-Bello}, {Aloisi}, {Macci{\`o}}, {Gallego-Laborda}, {Brodie}, \& {Merrifield}}]{Martinez-Delgado2012}
{Mart{\'{\i}}nez-Delgado}, D., {Romanowsky}, A.~J., {Gabany}, R.~J., {et~al.} 2012, \apjl, 748, L24

\bibitem[{{McQuinn} {et~al.}(2010){McQuinn}, {Skillman}, {Cannon}, {Dalcanton}, {Dolphin}, {Hidalgo-Rodr{\'\i}guez}, {Holtzman}, {Stark}, {Weisz}, \& {Williams}}]{McQuinn2010}
{McQuinn}, K. B.~W., {Skillman}, E.~D., {Cannon}, J.~M., {et~al.} 2010, \apj, 724, 49

\bibitem[{{Nidever} {et~al.}(2017){Nidever}, {Olsen}, {Walker}, {Vivas}, {Blum}, {Kaleida}, {Choi}, {Conn}, {Gruendl}, {Bell}, {Besla}, {Mu{\~n}oz}, {Gallart}, {Martin}, {Olszewski}, {Saha}, {Monachesi}, {Monelli}, {de Boer}, {Johnson}, {Zaritsky}, {Stringfellow}, {van der Marel}, {Cioni}, {Jin}, {Majewski}, {Martinez-Delgado}, {Monteagudo}, {No{\"e}l}, {Bernard}, {Kunder}, {Chu}, {Bell}, {Santana}, {Frechem}, {Medina}, {Parkash}, {Navarrete}, \& {Hayes}}]{Nidever2017}
{Nidever}, D.~L., {Olsen}, K., {Walker}, A.~R., {et~al.} 2017, \aj, 154, 199

\bibitem[{{Ott} {et~al.}(2012){Ott}, {Stilp}, {Warren}, {Skillman}, {Dalcanton}, {Walter}, {de Blok}, {Koribalski}, \& {West}}]{Ott2012}
{Ott}, J., {Stilp}, A.~M., {Warren}, S.~R., {et~al.} 2012, \aj, 144, 123

\bibitem[{{Pascale} {et~al.}(2022){Pascale}, {Annibali}, {Tosi}, {Marinacci}, {Nipoti}, {Bellazzini}, {Romano}, {Sacchi}, {Aloisi}, \& {Cignoni}}]{Pascale2022}
{Pascale}, R., {Annibali}, F., {Tosi}, M., {et~al.} 2022, \mnras, 509, 2940

\bibitem[{{Pascale} {et~al.}(2024){Pascale}, {Annibali}, {Tosi}, {Nipoti}, {Marinacci}, {Bellazzini}, {Cannon}, {Schisgal}, \& {Calura}}]{Pascale2024}
{Pascale}, R., {Annibali}, F., {Tosi}, M., {et~al.} 2024, arXiv e-prints, arXiv:2405.12284

\bibitem[{{Pascale} {et~al.}(2021){Pascale}, {Bellazzini}, {Tosi}, {Annibali}, {Marinacci}, \& {Nipoti}}]{Pascale2021}
{Pascale}, R., {Bellazzini}, M., {Tosi}, M., {et~al.} 2021, \mnras, 501, 2091

\bibitem[{{Paudel} {et~al.}(2018){Paudel}, {Smith}, {Yoon}, {Calder{\'o}n-Castillo}, \& {Duc}}]{Paudel2018}
{Paudel}, S., {Smith}, R., {Yoon}, S.~J., {Calder{\'o}n-Castillo}, P., \& {Duc}, P.-A. 2018, \apjs, 237, 36

\bibitem[{{Planck Collaboration} {et~al.}(2016){Planck Collaboration}, {Ade}, {Aghanim}, {Arnaud}, {Ashdown}, {Aumont}, {Baccigalupi}, {Banday}, {Barreiro}, {Bartlett}, {Bartolo}, {Battaner}, {Battye}, {Benabed}, {Beno{\^\i}t}, {Benoit-L{\'e}vy}, {Bernard}, {Bersanelli}, {Bielewicz}, {Bock}, {Bonaldi}, {Bonavera}, {Bond}, {Borrill}, {Bouchet}, {Boulanger}, {Bucher}, {Burigana}, {Butler}, {Calabrese}, {Cardoso}, {Catalano}, {Challinor}, {Chamballu}, {Chary}, {Chiang}, {Chluba}, {Christensen}, {Church}, {Clements}, {Colombi}, {Colombo}, {Combet}, {Coulais}, {Crill}, {Curto}, {Cuttaia}, {Danese}, {Davies}, {Davis}, {de Bernardis}, {de Rosa}, {de Zotti}, {Delabrouille}, {D{\'e}sert}, {Di Valentino}, {Dickinson}, {Diego}, {Dolag}, {Dole}, {Donzelli}, {Dor{\'e}}, {Douspis}, {Ducout}, {Dunkley}, {Dupac}, {Efstathiou}, {Elsner}, {En{\ss}lin}, {Eriksen}, {Farhang}, {Fergusson}, {Finelli}, {Forni}, {Frailis}, {Fraisse}, {Franceschi}, {Frejsel}, {Galeotta}, {Galli}, {Ganga}, {Gauthier}, {Gerbino}, {Ghosh}, {Giard},
  {Giraud-H{\'e}raud}, {Giusarma}, {Gjerl{\o}w}, {Gonz{\'a}lez-Nuevo}, {G{\'o}rski}, {Gratton}, {Gregorio}, {Gruppuso}, {Gudmundsson}, {Hamann}, {Hansen}, {Hanson}, {Harrison}, {Helou}, {Henrot-Versill{\'e}}, {Hern{\'a}ndez-Monteagudo}, {Herranz}, {Hildebrandt}, {Hivon}, {Hobson}, {Holmes}, {Hornstrup}, {Hovest}, {Huang}, {Huffenberger}, {Hurier}, {Jaffe}, {Jaffe}, {Jones}, {Juvela}, {Keih{\"a}nen}, {Keskitalo}, {Kisner}, {Kneissl}, {Knoche}, {Knox}, {Kunz}, {Kurki-Suonio}, {Lagache}, {L{\"a}hteenm{\"a}ki}, {Lamarre}, {Lasenby}, {Lattanzi}, {Lawrence}, {Leahy}, {Leonardi}, {Lesgourgues}, {Levrier}, {Lewis}, {Liguori}, {Lilje}, {Linden-V{\o}rnle}, {L{\'o}pez-Caniego}, {Lubin}, {Mac{\'\i}as-P{\'e}rez}, {Maggio}, {Maino}, {Mandolesi}, {Mangilli}, {Marchini}, {Maris}, {Martin}, {Martinelli}, {Mart{\'\i}nez-Gonz{\'a}lez}, {Masi}, {Matarrese}, {McGehee}, {Meinhold}, {Melchiorri}, {Melin}, {Mendes}, {Mennella}, {Migliaccio}, {Millea}, {Mitra}, {Miville-Desch{\^e}nes}, {Moneti}, {Montier}, {Morgante}, {Mortlock},
  {Moss}, {Munshi}, {Murphy}, {Naselsky}, {Nati}, {Natoli}, {Netterfield}, {N{\o}rgaard-Nielsen}, {Noviello}, {Novikov}, {Novikov}, {Oxborrow}, {Paci}, {Pagano}, {Pajot}, {Paladini}, {Paoletti}, {Partridge}, {Pasian}, {Patanchon}, {Pearson}, {Perdereau}, {Perotto}, {Perrotta}, {Pettorino}, {Piacentini}, {Piat}, {Pierpaoli}, {Pietrobon}, {Plaszczynski}, {Pointecouteau}, {Polenta}, {Popa}, {Pratt}, {Pr{\'e}zeau}, {Prunet}, {Puget}, {Rachen}, {Reach}, {Rebolo}, {Reinecke}, {Remazeilles}, {Renault}, {Renzi}, {Ristorcelli}, {Rocha}, {Rosset}, {Rossetti}, {Roudier}, {Rouill{\'e} d'Orfeuil}, {Rowan-Robinson}, {Rubi{\~n}o-Mart{\'\i}n}, {Rusholme}, {Said}, {Salvatelli}, {Salvati}, {Sandri}, {Santos}, {Savelainen}, {Savini}, {Scott}, {Seiffert}, {Serra}, {Shellard}, {Spencer}, {Spinelli}, {Stolyarov}, {Stompor}, {Sudiwala}, {Sunyaev}, {Sutton}, {Suur-Uski}, {Sygnet}, {Tauber}, {Terenzi}, {Toffolatti}, {Tomasi}, {Tristram}, {Trombetti}, {Tucci}, {Tuovinen}, {T{\"u}rler}, {Umana}, {Valenziano}, {Valiviita}, {Van Tent},
  {Vielva}, {Villa}, {Wade}, {Wandelt}, {Wehus}, {White}, {White}, {Wilkinson}, {Yvon}, {Zacchei}, \& {Zonca}}]{Planck2016}
{Planck Collaboration}, {Ade}, P.~A.~R., {Aghanim}, N., {et~al.} 2016, \aap, 594, A13

\bibitem[{{Schulte-Ladbeck} {et~al.}(2000){Schulte-Ladbeck}, {Hopp}, {Greggio}, \& {Crone}}]{Schulte2000}
{Schulte-Ladbeck}, R.~E., {Hopp}, U., {Greggio}, L., \& {Crone}, M.~M. 2000, \aj, 120, 1713

\bibitem[{{Starkenburg} {et~al.}(2016{\natexlab{a}}){Starkenburg}, {Helmi}, \& {Sales}}]{Starkenburg2016a}
{Starkenburg}, T.~K., {Helmi}, A., \& {Sales}, L.~V. 2016{\natexlab{a}}, \aap, 587, A24

\bibitem[{{Starkenburg} {et~al.}(2016{\natexlab{b}}){Starkenburg}, {Helmi}, \& {Sales}}]{Starkenburg2016b}
{Starkenburg}, T.~K., {Helmi}, A., \& {Sales}, L.~V. 2016{\natexlab{b}}, \aap, 595, A56

\bibitem[{{Stierwalt} {et~al.}(2015){Stierwalt}, {Besla}, {Patton}, {Johnson}, {Kallivayalil}, {Putman}, {Privon}, \& {Ross}}]{Stierwalt2015}
{Stierwalt}, S., {Besla}, G., {Patton}, D., {et~al.} 2015, \apj, 805, 2

\bibitem[{{Torrealba} {et~al.}(2018){Torrealba}, {Belokurov}, {Koposov}, {Bechtol}, {Drlica-Wagner}, {Olsen}, {Vivas}, {Yanny}, {Jethwa}, {Walker}, {Li}, {Allam}, {Conn}, {Gallart}, {Gruendl}, {James}, {Johnson}, {Kuehn}, {Kuropatkin}, {Martin}, {Martinez-Delgado}, {Nidever}, {No{\"e}l}, {Simon}, {Stringfellow}, \& {Tucker}}]{Torrealba2018}
{Torrealba}, G., {Belokurov}, V., {Koposov}, S.~E., {et~al.} 2018, \mnras, 475, 5085

\bibitem[{{Vera-Casanova} {et~al.}(2022){Vera-Casanova}, {G{\'o}mez}, {Monachesi}, {Gargiulo}, {Pallero}, {Grand}, {Marinacci}, {Pakmor}, {Simpson}, {Frenk}, \& {Morales}}]{Vera2022}
{Vera-Casanova}, A., {G{\'o}mez}, F.~A., {Monachesi}, A., {et~al.} 2022, \mnras, 514, 4898

\bibitem[{{Wang} {et~al.}(2020){Wang}, {Bose}, {Frenk}, {Gao}, {Jenkins}, {Springel}, \& {White}}]{Wang2020}
{Wang}, J., {Bose}, S., {Frenk}, C.~S., {et~al.} 2020, \nat, 585, 39

\bibitem[{{Weisz} {et~al.}(2011){Weisz}, {Dalcanton}, {Williams}, {Gilbert}, {Skillman}, {Seth}, {Dolphin}, {McQuinn}, {Gogarten}, {Holtzman}, {Rosema}, {Cole}, {Karachentsev}, \& {Zaritsky}}]{Weisz2011}
{Weisz}, D.~R., {Dalcanton}, J.~J., {Williams}, B.~F., {et~al.} 2011, \apj, 739, 5

\bibitem[{{Wetzel} {et~al.}(2016){Wetzel}, {Hopkins}, {Kim}, {Faucher-Gigu{\`e}re}, {Kere{\v{s}}}, \& {Quataert}}]{Wetzel2016}
{Wetzel}, A.~R., {Hopkins}, P.~F., {Kim}, J.-h., {et~al.} 2016, \apjl, 827, L23

\bibitem[{{White} \& {Rees}(1978)}]{White78}
{White}, S.~D.~M. \& {Rees}, M.~J. 1978, \mnras, 183, 341

\bibitem[{{Zhang} {et~al.}(2020){Zhang}, {Smith}, {Oh}, {Paudel}, {Duc}, {Boselli}, {C{\^o}t{\'e}}, {Ferrarese}, {Gao}, {Hunter}, {Puzia}, {Peng}, {Rong}, {Shin}, \& {Zhao}}]{Hong2020}
{Zhang}, H.-X., {Smith}, R., {Oh}, S.-H., {et~al.} 2020, \apj, 900, 152

\bibitem[{{Zheng} {et~al.}(2024){Zheng}, {Faerman}, {Oppenheimer}, {Putman}, {McQuinn}, {Kirby}, {Burchett}, {Telford}, {Werk}, \& {Kim}}]{Zheng2024}
{Zheng}, Y., {Faerman}, Y., {Oppenheimer}, B.~D., {et~al.} 2024, \apj, 960, 55

\end{thebibliography}

\begin{appendix}
\section{Stellar populations in the substructures}
\label{app:stars}

\begin{figure*}
\centering
\includegraphics[width=\linewidth]{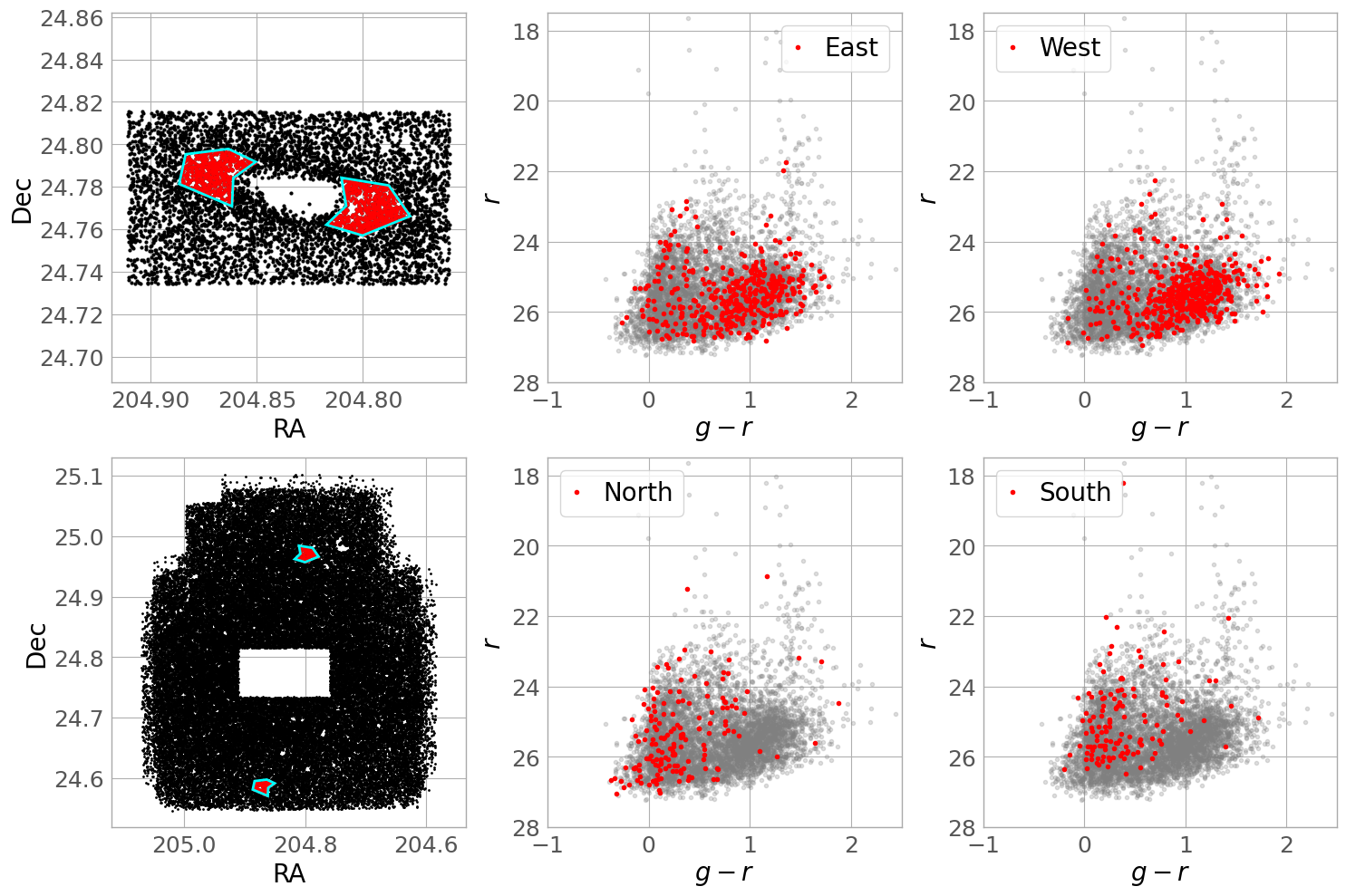}
\caption{\textit{Top row.} Spatial distribution of the central portion of our image for  UGC 8638 and CMDs of the stars contained in the two highlighted regions (cyan boxes); the gray background points are all the stars in this central field. \textit{Bottom row.} Map of the whole field (with an empty rectangle corresponding to the central area in the top panel, where the galaxy is); the background gray points are the same as before, as a reference, while the red points show the stars contained in the two control fields (cyan boxes).}
\label{fig:bkg}
\end{figure*}

As an additional check to our RGB selection, we compared the CMDs of all stars contained in each polygon highlighted in Figures \ref{fig:ngc5238} to \ref{fig:ugc8760} with those of similar control fields of the same area, but taken further away from the galaxies. Figure \ref{fig:bkg} shows an example of these different CMDs in UGC~8638. The top row contains the spatial map of the central portion of our image and the CMDs of the stars (mainly RGBs) contained in the two northeastern and southwestern boxes (the gray background points are all the stars in this central field). The bottom row shows the whole field (with an empty rectangle corresponding to the central area where the galaxy is, shown in the top row); the background gray points are the same as before, as a reference, while the red points show the stars contained in the two control fields. The excess of RGB stars in the areas containing the tidal features is striking, while in the control fields we find mostly background blue sequence galaxies \citep{Blanton2009} and probably a few residual MW disc stars, once more confirming that the over-densities shown in the RGB maps are statistically significant.

\end{appendix}

\end{document}